\documentclass[review,authoryear,10pt]{elsarticle}
\usepackage{amsmath}
\usepackage{array}
\usepackage{calc}
\usepackage{amssymb}
\usepackage{natbib}
\graphicspath{{./Figs/}} 
\usepackage{subfigure}
\usepackage[normalem]{ulem}
\usepackage{tabularx}
\usepackage{graphicx}
\usepackage{multirow}
\usepackage{ctable}
\usepackage{soul}
\begin{document}

\begin{frontmatter}
\title{Filtered models for reacting gas-particle flows}
\author[prt]{William Holloway\corref{cor1}}
\ead{whollowa@princeton.edu}
\author[prt]{Sankaran Sundaresan}
\ead{sundar@princeton.edu}

\address[prt] {Department of Chemical and Biological Engineering, Princeton University, Princeton, NJ 08544}
\cortext[cor1]{Corresponding Author}

\begin{abstract} 
Using the kinetic-theory-based two-fluid models as a starting point, we develop \emph{filtered} two-fluid models for a gas-particle flow in the presence of an isothermal, first-order, solid-catalyzed reaction of a gaseous species.  As a consequence of the \emph{filtering} procedure, terms describing the \emph{filtered} reaction rate and \emph{filtered} reactant dispersion need to be constituted in order to close the \emph{filtered} species balance equation.  In this work, a constitutive relation for \emph{filtered} reaction rate is developed by performing fine-grid, two-fluid model simulations of an isothermal, solid-catalyzed, first-order reaction in a periodic domain.  It is observed that the \emph{cluster-scale} effectiveness factor, defined as the ratio between the reaction rate observed in a fine-grid simulation to that observed in a coarse-grid simulation, can be substantially smaller than unity, and it manifests an inverted bell shape dependence on \emph{filtered} particle volume fraction in all simulation cases.  Moreover, the magnitude of the deviation in the \emph{cluster-scale} effectiveness factor from unity is a strong function of the \emph{meso-scale} Thiele modulus and dimensionless filter size. Thus coarse-grid simulations of a reacting gas-particle flow will overestimate the reaction rate if the \emph{cluster-scale} effectiveness factor is not accounted for.  
\end{abstract}


\end{frontmatter}


\section{Introduction}

Gas-particle flows are commonly encountered in the chemical and energy conversion industries in the form of fluidized beds, risers, and other pneumatic conveying units.  The flow behaviors of gas-particle systems are commonly analyzed using continuum models that treat the particle and fluid phases as interpenetrating continua~\citep{Gidaspow1994,Fan1998,Jackson2000}.  Finite-volume simulations of these continuum, or `two-fluid' models, reveal that the scale of the gas-particle flow structures that are observed is a strong function of the grid size used in the simulation, with grid size independent results being obtained when the ratio between the grid size and particle size is $\emph{O}(10)$~\citep{Agrawal2001,Andrews2005,Igci2008}.  Grid cell sizes of ten particle diameters are not computationally affordable when performing two-fluid model simulations of large fluidized bed reactors that are often tens of meters tall and many meters in cross-section.  As a result, coarse-grid simulations of large fluidized beds are common practice.  However, these coarse-grid simulations effectively neglect the presence of fine scale gas-particle flow structures that are manifested by the two-fluid model.  To enable accurate coarse-grid simulation of the two-fluid model equations one must account for the effect of the fine-scale gas-particle flow structures in the coarse-grid simulation~\citep{Andrews2005,Igci2008}.  This approach is embodied by the recent development of a \emph{filtered} two-fluid model approach for non-reacting gas-particle flows where effective closures for the \emph{meso-scale} fluid-particle drag force, particle phase stress, and particle phase viscosity were extracted from fine-grid simulations of the two-fluid model equations~\citep{Igci2008,Igci2011b,Igci2011,Igci2011a,parmentier2011}.  While \emph{filtered} two-fluid models for non-reacting monodisperse gas-particle flows have been developed, verified against fine-grid simulations, and validated against experimental observations~\citep{Igci2011b,Igci2011,Igci2011a}, the extension of these models to reacting and polydisperse gas-particle flows remains an open problem.

Fine-grid two-fluid model simulations of reacting gas-particle flow have been shown to yield quantitative agreement with experimental observations of conversion in solid-catalyzed ozone decomposition processes when grid-independent solutions are obtained~\citep{Syamlal2003}.  However, due to the computational expense of performing grid-independent simulations, coarse numerical resolution is often used to simulate reacting gas-particle flows in the continuum model framework.  By not accounting for the fine-scale structure these coarse-grid simulations have been shown to over-predict the conversion of ozone in a bubbling fluidized bed~\citep{Zimmermann2005}.  In addition, a few studies have also demonstrated that small-scale gas-particle flow structures that form in devices like fluidized beds are responsible for the wide variability in mass transfer models that is present within the fluidization literature~\citep{Dong2008,Dong2008a,Kashyap2010,Kashyap2011}.  

The primary objective of this work is to demonstrate the need for \emph{filtered} models for reacting gas-particle flows by performing fine-grid simulations of a first-order, isothermal, solid catalyzed reaction in a periodic domain and filtering the results.  It will be shown that the presence of particle clustering in fine-grid simulations leads to an effective reaction rate that is substantially smaller than what would be predicted via coarse-grid simulation.  We define the ratio of the reaction rate in the fine-grid simulation to that in the coarse-grid simulation as the \emph{cluster-scale} effectiveness factor.  It will be shown that this \emph{cluster-scale} effectiveness factor is a strong function of dimensionless filter size $\hat{\Delta}=\Delta |\boldsymbol{g}|/v_t^2$, and other model parameters.  Here, $\Delta$ is the filter size, $\boldsymbol{g}$ is the gravitational acceleration vector, and $v_t$ is the terminal settling velocity of an isolated particle.

Finally, it is shown that grid resolutions finer than those used to deduce \emph{filtered} models for non-reacting gas-particle flows are necessary in order to obtain grid-independent \emph{filtered} closures for reacting gas-particle flows.  This observation is supported by a recent work by~\cite{Cloete2011}.  Due to this grid size dependence we propose a \emph{filtered} reaction rate model based upon extrapolated effectiveness factor data obtained from fine-grid simulations. 

\section{Microscopic two-fluid model equations}

The two-fluid model consists of balance equations for gas- and particle-phase mass and momentum with an additional transport equation governing evolution of the particle-phase granular temperature.  The evolution equations that constitute this Eulerian framework are presented in Table 1, in addition to the constitutive relations that close the balance equations.  The constitutive relations given in Table 1 were chosen to be consistent with earlier studies in filtered model development within our research group~\citep{Agrawal2001,Andrews2005,Igci2008,Igci2011,Igci2011a,Igci2011b}.  Here we note that while we have restricted our attention to a specific set of constitutive relations, it has been shown that the clustering and bubbling phenomena manifested by the two-fluid model framework are robust to changes in constitutive relations~\citep{Glasser1998}.  However, we do expect that the constitutive relations comprising the microscopic two-fluid models will have quantitative effects on the filtered models that arise from fine-grid simulations.  Therefore, while we have employed a certain set of constitutive relations in this study, filtered models can be developed from fine-grid, two-fluid model simulations using a different set of constitutive relations as well.  

The effective diffusion coefficient $D^*$ and effective gas-phase viscosity $\mu_g^*$ appearing in eqs.~(\ref{eq:cont_species}) and~(\ref{eq:gas_stress}) will differ from the corresponding molecular properties as they also account for the enhanced scalar and momentum transport that occurs due to additional \emph{pseudo-turbulent} transport occurring as a result of microscopic interactions between individual particles and the fluid.  However, two-fluid model simulations have revealed that the solutions manifested by the two-fluid model are insensitive to the value of $\mu_g^*$~\citep{Agrawal2001}.  Therefore, we set $\mu_g^*$ equal to the molecular viscosity $\mu_g$.  In addition, the effective diffusion coefficient $D^*$ is kept on the order of the bulk molecular diffusivity $D$, which is generally $O(10^{-5}) \ m^2/sec$ for many different gas species.  
 
Here we investigate the role of clustering on a model, isothermal, solid-catalyzed, first-order chemical reaction
\begin{equation}
A \rightarrow B \qquad R_i=-k_{\text{eff}} \phi \rho_g \chi_{g_{A}}
\label{eq:rate_law}
\end{equation}
where $\phi$ is the particle volume fraction, and $\chi_{g_{A}}$ is the mass fraction of species $A$ in the gas phase.  For first order reaction kinetics, the effective reaction rate constant $k_{\text{eff}}$ based on the bulk concentration $\chi_{g_{i}}$ is related to the intrinsic reaction rate constant $k$ via
\begin{equation}
k_\text{eff}=\frac{\eta_i k}{1+\Phi^2 \eta_i/Bi} \quad \eta_i=\frac{1}{\Phi}\left(\frac{1}{\tanh(3\Phi)}-\frac{1}{(3\Phi)}\right),
\label{eq:micro_eta}
\end{equation}
where $\eta_i$ is the \emph{intra-particle} effectiveness in the absence of mass transport limitations, $\Phi=\sqrt{k a/D_I}$ is the Thiele modulus, $a$ is the volume to surface area ratio, and $Bi$ is the Biot number for mass transport given as $Bi=k_m a/D_I$~\citep{Rawlings2009}.  Here, $k_m$ is the convective mass transport coefficient, $a$ is the particle radius, and $D_I$ is the \emph{intra-particle} diffusivity.  It is important to note that the effective reaction rate constants used in two-fluid model simulations incorporate the combined effect of mass transport resistance and \emph{intra-particle} effects as well.  In this study all gas-solid flow computations are performed for FCC catalyst particles fluidized by air $-$ the physical properties for solid and fluid phases are given in Table 2.

\begin{table}
\begin{tabularx}{5 in}{X}
\cline{1-1}
\hline
\vspace{0.1 in}
\uline{\textbf{Evolution Equations}} 
\begin{equation}
\frac{\partial \left(\rho_{s}\phi\right)}{\partial t}+\nabla \cdot \left(\rho_s \phi \boldsymbol{v}\right)=0
\end{equation}
\begin{equation}
\frac{\partial \left(\rho_g \epsilon_g \right)}{\partial t}+\nabla \cdot \left(\rho_g \epsilon_g \boldsymbol{u}\right)=0
\label{eq:cont_fluid}
\end{equation} 
\begin{equation}
\frac{\partial \left(\rho_g \epsilon_g \chi_{g_{i}} \right)}{\partial t}+\nabla \cdot \left(\rho_g \epsilon_g \chi_{g_{i}}  \boldsymbol{u}\right)=-\nabla\cdot \left(D^* \epsilon_g \nabla \chi_{g_{i}}\right)+R_{i}
\label{eq:cont_species}
\end{equation} 
\begin{equation}
\frac{\partial \left(\rho_s \phi \boldsymbol{v} \right)}{\partial t}+\nabla \cdot \left(\rho_s \phi \boldsymbol{v}  \boldsymbol{v}\right)=-\phi\nabla\cdot \boldsymbol{\sigma}_g- \nabla\cdot \boldsymbol{\sigma}_s+\boldsymbol{f_{D}}+\rho_s\phi\boldsymbol{g}
\label{eq:mom_part}
\end{equation}
\begin{equation}
\frac{\partial \left(\rho_g \epsilon_g \boldsymbol{u} \right)}{\partial t}+\nabla \cdot \left(\rho_g \epsilon_g \boldsymbol{u}  \boldsymbol{u}\right)=-\epsilon_g\nabla\cdot \boldsymbol{\sigma}_g -\boldsymbol{f_{D}}+\rho_g\epsilon_g\boldsymbol{g}
\label{eq:mom_fluid}
\end{equation} 
\begin{equation}
\frac{3}{2}\left[\frac{\partial \left(\rho_s \phi T\right)}{\partial t} +\nabla \cdot \left(\rho_s \phi T \boldsymbol{v}  \right)\right]=-\nabla\cdot \boldsymbol{q} -\boldsymbol{\sigma}_s:\nabla \boldsymbol{v}-J_\text{vis}-J_\text{coll}+\Gamma_\text{slip}
\label{eq:gran_temp}
\end{equation} 
\uline{\textbf{Closures}} \\
\uline{ Gas phase stress tensor} $\left(\boldsymbol{\sigma}_g\right)$ 
\begin{equation}
\boldsymbol{\sigma}_g=p_g\boldsymbol{I}-\mu_g^*\left(\nabla \boldsymbol{u} +\left(\nabla \boldsymbol{u}\right)^T-\frac{2}{3}\left(\nabla \cdot \boldsymbol{u}\right)\boldsymbol{I}\right)
\label{eq:gas_stress}
\end{equation}
\uline{Solid phase stress tensor} $\left(\boldsymbol{\sigma}_s\right)$ 
\begin{equation}
\boldsymbol{\sigma}_s=\left(\rho_s\phi\left(1+4\phi g_0\right)T- \eta \mu_b\left(\nabla \cdot \boldsymbol{u}\right)\right)\boldsymbol{I}-2\mu_s \boldsymbol{S},\qquad \eta=\frac{1+e_p}{2},
\label{eq:solid_stress}
\end{equation}
\begin{equation}
g_0=\frac{1}{1-\left(\phi/\phi_{max}\right)^{1/3}}, \qquad \mu_b=\frac{256 \mu \phi^2 g_0}{5\pi}, \qquad \mu=\frac{5 \rho_s d\sqrt{\pi T}}{96},
\label{eq:rad_dist_mu_b}
\end{equation}
\begin{equation}
\mu_s=\frac{1.2 \mu^*}{g_0\eta\left(2-\eta\right)}\left(1+\frac{8}{5}\phi\eta g_0\right)\left(1+\frac{8}{5}\eta\left(3\eta-2\right)\phi g_0+\frac{6}{5}\eta\mu_b\right),
\label{eq:mu_s}
\end{equation}
\begin{equation}
\mu^*=\frac{\mu}{1+\frac{2\beta \mu}{\left(\rho_s \phi\right)^2 g_0 T}}, \qquad  \boldsymbol{S}=\frac{1}{2}\left(\nabla \boldsymbol{v}+\left(\nabla \boldsymbol{v}\right)^T\right)-\frac{1}{3}\left(\nabla \cdot \boldsymbol{v}\right)\boldsymbol{I}.
\label{eq:mu_star}
\end{equation}
\uline{Fluid-particle drag force} $\left(\boldsymbol{f_D}\right)$
\begin{equation}
\boldsymbol{f_D}=\beta\left(\boldsymbol{u}-\boldsymbol{v}\right), \quad \beta=\frac{C_D}{d^2} Re_g \phi \mu_g^* \epsilon_g^{-2.65} , \quad Re_g=\frac{\epsilon_g \rho_g d |\boldsymbol{u}-\boldsymbol{v}|}{\mu_g^*}
\label{eq:drag_force}
\end{equation}
\begin{equation}
 C_D =\left\{ \begin{array}{l r}
         18\left(Re_g^{-1}+0.15Re_g^{-0.313}\right) & \quad\quad Re_g<1000 \\  
         0.33 & \quad\quad Re_g \geq 1000
         \end{array}\right.
\label{eq:C_D}
\end{equation}
\end{tabularx}
\end{table}
\begin{table}
\begin{tabularx}{5 in}{X}
\cline{1-1}
\hline
\vspace{0.1 in}
\uline{Diffusive flux of granular energy} $\left(\boldsymbol{q}\right)$
\begin{equation}
\boldsymbol{q}=-\lambda\nabla T
\label{eq:heat_flux}
\end{equation}
\begin{equation} 
\lambda=\frac{\lambda^*}{g_0}\left(\left(1+\frac{12}{5}\eta \phi g_0\right)\left(1+\frac{12}{5}\eta^2\left(4\eta-3\right)\phi g_0\right)+\frac{64}{25 \pi}\left(41-33\eta\right)\eta^2\phi^2g_0^2\right) 
\label{eq:conduct}
\end{equation}
\begin{equation} 
\lambda^*=\frac{\lambda_i}{1+\frac{6\beta \lambda_i}{5\left(\rho_s \phi\right)^2 g_0 T}}; \qquad \lambda_i=\frac{75 \rho_s d \sqrt{\pi T}}{48 \eta \left(41-33\eta \right)}
\end{equation}\uline{Dissipation of granular energy via inelastic collisions} $\left(J_\text{coll}\right)$
\begin{equation} 
J_\text{coll}=\frac{48}{\sqrt{\pi}}\eta\left(1-\eta\right)\frac{\rho_s \phi^2}{d} g_0T^{3/2}
\end{equation}
\uline{Dissipation of granular energy via viscous action of the fluid phase} $\left(J_\text{vis}\right)$
\begin{equation} 
J_\text{vis}=\frac{54 \phi \mu_g^* T}{d^2}R_{\mathrm{diss}}
\end{equation}
\begin{equation}
R_{\mathrm{diss}}=1+3\sqrt{\frac{\phi}{2}}+\frac{135}{64}\phi \mathrm{ln} (\phi)+11.26 \phi\left(1-5.1\phi+16.57 \phi^2-21.77\phi^3\right) -\phi g_0\ln(0.01)
\end{equation}
\uline{Production of granular energy through slip between phases} $\left(\Gamma_\mathrm{slip}\right)$
\begin{equation} 
\Gamma_\text{slip}=\frac{81 \phi \mu_g^{*2}|\boldsymbol{u}-\boldsymbol{v}|^2}{g_0 d^3\rho_g\sqrt{\pi T}}\frac{R_{drag}^2}{1+3.5\sqrt{\phi}+5.9\phi}
\end{equation}
\begin{equation}
 R_{drag} =\left\{ \begin{array}{l r}
         \frac{1+3\sqrt{\phi/2}+(135/64)\phi \ln(\phi)+17.14 \phi}{1+0.681\phi-8.48\phi^2+8.16\phi^3} & \quad\quad \phi<0.40 \\  
         \frac{10\phi}{\left(1-\phi\right)^3}+0.7 & \quad\quad \phi \geq 0.40 \\
         \end{array}
         \right.
\label{eq:R_drag}
\end{equation}
\end{tabularx}
\caption{Microscropic two-fluid model equations for an isothermal, reacting gas-particle flow.  Here, $\rho_s$ and $\rho_g$ are particle and fluid phase densities respectively; $\phi$ is the particle volume fraction; $\epsilon_g$ is the gas void fraction; $\boldsymbol{v}$ and $\boldsymbol{u}$ are particle and fluid phase velocities; $R_i$ is the rate of production of species \emph{i}; $D^*$ is the effective diffusion coefficient; $\mu_g^*$ and $\mu_s$ are the effective gas phase viscosity and solid phase viscosity, respectively; $g_0$ is the radial distribution function at contact; $e_p$ is the coefficient of restitution; $\lambda$ is the conductivity of pseudothermal energy; and $T$ is the granular temperature}
\end{table}

\begin{table}
\begin{center}
\begin{tabularx}{6cm}{p{2cm}X}
\hline
$d$ & $7.5\times 10^{-5}$ \ m \\
$\rho_s$ & $1500$ \ $kg/m^3$ \\
$\rho_g$ & $1.3$  \ $kg/m^3$ \\
$\mu_g$ & $1.8\times 10^{-5}$ \ $kg/\left( m \ s\right)$ \\
$e_p$ & $0.9$ \\
\hline
\end{tabularx}
\caption{List of physical parameters for solid and fluid phases.}
\end{center}
\end{table}

\section{\emph{Filtered} two-fluid model equations}
\label{sec:filt_two_fluid}
The \emph{filtered} two-fluid models are obtained by performing a spatial average of the microscopic two-fluid model equations.  The consequence of the filtering approach is that the fine-scale gas-particle flow structure that occurs on a length scale smaller than the filter size is accounted for through residual terms that must be constituted from theoretical considerations or fine-grid two-fluid model simulation results.  In the case of non-reacting monodisperse gas-particle flows, \emph{filtered} models with accompanying constitutive relations for residual terms have been shown to yield quantitatively similar macroscopic behaviors to those observed in fine-grid simulations of the same flow problem, with a dramatic savings in computational time~\citep{Igci2011}.  Here we follow, the filtering procedure given in~\cite{Igci2008}.

The particle volume fraction and gas-species mass fractions obtained from fine-grid simulations can be represented as $\phi(\boldsymbol{y},t)$, where $\boldsymbol{y}$ and \emph{t} represent the location and time variables, respectively.  The \emph{filtered} particle volume fraction $\left< \phi\right>(\boldsymbol{x_0},t)$ is given as 
\begin{equation}
\left< \phi\right>(\boldsymbol{x_0},t) =\int_{V_{\infty}} G(\boldsymbol{x_0},\boldsymbol{y})\phi(\boldsymbol{y},t) d\boldsymbol{y}
\end{equation}
where $G(\boldsymbol{x_0},\boldsymbol{y})$ is a weight function, and $\boldsymbol{x_0}$ is the spatial location of the filter center.  In this work $\left<\cdot\right>$ will be used to indicate filtered variables.  We require that $\int_{V_{\infty}}G(\boldsymbol{x_0},\boldsymbol{y})d\boldsymbol{y}=1$, and in this study we use a top hat filter for $G(\boldsymbol{x_0},\boldsymbol{y})$.  \emph{Filtered} gas-species mass fraction $\left< \chi_{g_{i}}\right>(\boldsymbol{x_0},t)$, particle phase velocity $\left<\boldsymbol{v}\right>(\boldsymbol{x_0},t)$,  and fluid phase velocity $\left< \boldsymbol{u}\right>(\boldsymbol{x_0},t)$ are given as
\begin{equation}
\left< \epsilon_g\right>(\boldsymbol{x_0},t)\left< \chi_{g_{i}}\right>(\boldsymbol{x_0},t)=\int_{V_{\infty}} G(\boldsymbol{x_0},\boldsymbol{y})\epsilon_g(\boldsymbol{y},t)\chi_{g_{i}}(\boldsymbol{y},t)d\boldsymbol{y}
\label{eq:filt_chi}
\end{equation}
\begin{equation}
\left< \phi\right>(\boldsymbol{x_0},t)\left< \boldsymbol{v}\right>(\boldsymbol{x_0},t)=\int_{V_{\infty}} G(\boldsymbol{x_0},\boldsymbol{y})\phi(\boldsymbol{y},t)\boldsymbol{v}(\boldsymbol{y},t)d\boldsymbol{y}
\label{eq:filt_vel_s}
\end{equation}
\begin{equation}
\left< \epsilon_g\right>(\boldsymbol{x_0},t)\left< \boldsymbol{u}\right>(\boldsymbol{x_0},t)=\int_{V_{\infty}} G(\boldsymbol{x_0},\boldsymbol{y})\epsilon_g(\boldsymbol{y},t)\boldsymbol{u}(\boldsymbol{y},t)d\boldsymbol{y}.
\label{eq:filt_vel_g}
\end{equation}
Upon filtering the microscopic species balance equation for component $A$ we obtain the following \emph{filtered} species balance equation
\begin{equation}
\frac{\partial \left( \rho_g \left<\epsilon_g\right>\left<\chi_{g_{A}}\right>\right)}{\partial t}+\nabla \cdot \left(\rho_g\left<\epsilon_g\chi_{g_{A}}\boldsymbol{u}\right>\right)=-\nabla \cdot \left(D\left<\epsilon_g\nabla \chi_{g_{A}}\right>\right)-k_\text{eff}\rho_g\left<\phi \chi_{g_{A}}\right>.
\label{eq:filt_chi_eq}
\end{equation}
In eq.~(\ref{eq:filt_chi_eq}) we invoke the definition of filtered variables given in eqs.~(\ref{eq:filt_chi})--(\ref{eq:filt_vel_g}).  Due to the fact that the reaction we are considering in this work is isothermal and produces no volume change, the \emph{filtered} momentum balance equations remain unchanged from those derived in the work of \cite{Igci2008}, and as such, they will not be presented here for the sake of brevity.  In eq.~(\ref{eq:filt_chi_eq}), there are several terms that appear as \emph{filtered} products of microscopic variables that must be constituted to solve the \emph{filtered} two-fluid model equations.  To facilitate filtered model simulations we must decompose the products of microscopic variables into mean and fluctuating parts.  Here, we define fluctuating variables as follows
\begin{equation}
\chi'_{g_{A}}(\boldsymbol{y},t)=\chi_{g_{A}}(\boldsymbol{y},t)-\left<\chi_{g_{A}}\right>(\boldsymbol{y},t) \qquad \boldsymbol{u}'(\boldsymbol{y},t)=\boldsymbol{u}(\boldsymbol{y},t)-\left<\boldsymbol{u}\right>(\boldsymbol{y},t) 
\end{equation}
\begin{equation}
\phi'(\boldsymbol{y},t)=\phi(\boldsymbol{y},t)-\left<\phi\right>(\boldsymbol{y},t) 
\end{equation}
where $\chi'_{g_{A}}$, $\boldsymbol{u}'$, and $\phi'$ represent species $A$ mass fraction, gas velocity, and volume fraction fluctuations, respectively.  Inserting the decomposition of mean and fluctuating parts into eq.~(\ref{eq:filt_chi_eq}), we obtain the following filtered species balance equation

\begin{equation}
\begin{aligned}
\frac{\partial \left( \rho_g \left<\epsilon_g\right>\left<\chi_{g_{A}}\right>\right)}{\partial t}+\nabla \cdot \left(\rho_g\left<\epsilon_g\right> \left<\chi_{g_{A}}\right>\left<\boldsymbol{u}\right>\right)=&- \nabla \cdot \left( \rho_g\left<\epsilon_g\chi'_{g_{A}}\boldsymbol{u}'\right>+D\left<\epsilon_g\nabla \chi_{g_{A}}\right> \right) \\ &-k_\text{eff}\rho_g\left(\left<\phi\right> \left<\chi_{g_{A}}\right>+\left<\chi'_{g_{A}}\right>\right) .
\end{aligned}
\label{eq:filt_chi_eq2}
\end{equation}

The first term appearing on the right hand side of eq.~(\ref{eq:filt_chi_eq2}) can be modeled as a dispersive term given by the following constitutive equation
\begin{equation}
D_\text{eff}\nabla \left<\chi_{g_{i}}\right>= \left( \rho_g\left<\epsilon_g\chi'_{g_{A}}\boldsymbol{u}'\right>+D\left<\epsilon_g\nabla \chi_{g_{A}}\right> \right),
\end{equation}
where $D_\text{eff}$ is a dispersion coefficient.  Dispersion coefficients defined in this way have been presented in the research literature~\citep{Loezos2002}.  The final term on the right hand side of eq.~(\ref{eq:filt_chi_eq2}) represents a \emph{filtered} reaction rate that must also be constituted in terms of filtered variables alone.  A straightforward method for constituting this reaction rate is to define a \emph{cluster-scale} effectiveness factor $\eta_{\Delta}$ which is given as 
\begin{equation}
\eta_{\Delta}=\frac{filtered \ reaction \ rate}{homogeneous \ reaction \ rate}=\frac{\left<\phi \chi_{g_{i}}\right>}{\left<\phi\right>\left<\chi_{g_{i}}\right>}.
\label{eq:effect_fac}
\end{equation}
However, in this paper we present \emph{filtered} models and associated constitutive relations for the \emph{non-locally} corrected \emph{cluster-scale} effectiveness factor $\eta_{\Delta}'$ defined as 
\begin{equation}
\eta_{\Delta}'=\frac{\left<\phi \chi_{g_{i}}\right>-m_2\nabla\left<\phi\right>^T\cdot\nabla\left<\chi_{g_{i}}\right>}{\left<\phi\right>\left<\chi_{g_{i}}\right>},
\label{eq:effect_fac2}
\end{equation}
where $m_2=\Delta^2/12$.  The approach of removing \emph{non-local} effects in the determination of filtered quantities was recently advanced by~\cite{parmentier2011}, and we apply the same procedure here for determining $\eta_{\Delta}'$.  The \emph{non-locally} corrected effectiveness factor is constructed by removing the dominant gradient terms that contribute the evaluation of the filtered product $\left<\phi \chi_{g_{i}}\right>$, which is done to allow the construction of \emph{filtered} constitutive relations in terms of local \emph{filtered} variables alone.  We present a derivation of the \emph{non-locally} corrected effectiveness factor in~\ref{sec:nonl_deriv}.  It will be shown below that both $\eta_{\Delta}$ and $\eta_{\Delta}'$ depend on volume fraction in a qualitatively similar way, but differ quantitatively.  All data presented in future sections and resulting \emph{filtered} models will pertain to $\eta_{\Delta}'$, thus necessitating one to track the gradients in filtered species mass fraction and volume fraction to enable the use of the resulting filtered models in coarse-grid simulations of reacting gas-particle flow.

Dimensional analysis of the parameters governing gas-particle flow suggests that the \emph{cluster-scale} effectiveness factor is a function of five independent dimensionless quantities, which are 

\begin{equation}
\Pi_1=\sqrt{\frac{k_\text{eff} d^2}{D}}=\hat{\Phi} \quad \Pi_2=\frac{|\boldsymbol{g}|d}{v_t^2} \quad \Pi_3=\frac{\mu_g}{\rho_g D}=\hat{Sc} 
\end{equation}

\begin{equation}
\Pi_4=\frac{\rho_s}{\rho_g}\quad \Pi_5=\frac{|\boldsymbol{g}|\Delta}{v_t^2}=\hat{\Delta}.
\end{equation}
One can also readily define additional dimensionless parameters like $Re_g=\rho_g v_t/\mu_g$ and the coefficient of restitution $e_p$ using dimensional analysis.  However, earlier works of~\cite{Igci2008} have shown that \emph{filtered} quantities do not display any significant dependence on $Re_g$ or $e_p$, and as such, they were not included in the dimensional analysis presented here.  Moreover, we found that $\Pi_2$ and $\Pi_4$ had a negligible effect on the \emph{cluster-scale} effectiveness factor, while the \emph{meso-scale} Thiele modulus $\hat{\Phi}$, Schmidt number $\hat{Sc}$, and the dimensionless filter size $\hat{\Delta}$ substantially alter the \emph{cluster-scale} effectiveness factor.  Therefore, all work presented below will only interrogate the effect of $\hat{\Phi}$, $\hat{Sc}$, and $\hat{\Delta}$ on $\eta_{\Delta}'$.  Here the term \emph{meso-scale} Thiele modulus is used to distinguish it from the Thiele modulus based on \emph{intra-particle} diffusivity $D_I$ given in eq.~(\ref{eq:micro_eta}). 

\section{Numerical implementation}

When performing periodic domain simulations of reacting gas-particle flows, one is faced with the following dilemma:  due to the imposition of periodic boundary conditions, the concentration of any reactant undergoing an irreversible reaction within the periodic domain decay to zero.  In order to facilitate gas-particle flow simulations in periodic domains, an alternate simulation strategy must be developed.  For the special case of a first-order reaction, it can readily be shown that one can track the evolution of $\kappa_i=\chi_{g_{i}}/\left<\chi_{g_{i}}\right>$ rather than $\chi_{g_{i}}$ itself, and relate the \emph{filtered} value of $\kappa_i$ directly to the \emph{cluster-scale} effectiveness factor.  The benefit of tracking the evolution of $\kappa_i$ lies in the fact that it has a non-zero statistically steady value, even though $\chi_{g_{i}}$ will decay to zero.  

Consider taking the species balance equation given by eq.~(\ref{eq:cont_species}) with the reaction rate expression given by eq.~(\ref{eq:rate_law}) and averaging it over the entire periodic domain, resulting in the following equation
\begin{equation}
\rho_g\left<\epsilon_g\right>\frac{d\left<\chi_{g_{i}}\right>}{dt}=-\frac{1}{V}\int_{V}k_\text{eff}\rho_g \phi \chi_{g_{i}}dV,
\end{equation}
where $V$ is the volume of the periodic domain.  If we now define the variable $\kappa_i=\chi_{g_{i}}/\left<\chi_{g_{i}}\right>$ and plug $\kappa_i$ into the species balance equation given by eq.~(\ref{eq:cont_species}) we obtain the following evolution equation
\begin{equation}
\begin{aligned}
\frac{\partial \left(\rho_g \epsilon_g \kappa_i \right)}{\partial t}+\nabla \cdot \left(\rho_g \epsilon_g \kappa_i  \boldsymbol{u}\right)= & -\nabla\cdot \left(D \epsilon_g \nabla \kappa_i\right)-k_\text{eff}\rho_g \phi \kappa_i
\\ & +\frac{\kappa_i \epsilon_g}{\left<\epsilon_g\right>}\frac{1}{V}\int_{V}k_\text{eff}\rho_g \phi \kappa_i dV.
\end{aligned}
\label{eq:kappa_eq}
\end{equation}
Due to the fact that the reaction rate expressions in this problem are first order, we are able to track the evolution of $\kappa_i$ without considering the time progression of $\chi_{g_{i}}$ or $\left<\chi_{g_{i}}\right>$.  The presence of non-zero source and sink terms on the right hand side of eq.~(\ref{eq:kappa_eq}) force statistically steady value of $\kappa_i$ to be non-zero regardless of the reaction rate constant used.  Therefore, in all work presented in following sections for the evolution of $\kappa_i$ is solved, and the \emph{cluster-scale} effectiveness factor is redefined in terms of $\kappa_i$ as

\begin{equation}
\eta_{\Delta}'=\frac{\left<\phi \kappa_i\right>-m_2\left(\nabla \left<\phi\right>\right)^T\cdot \nabla \left<\kappa_i\right>}{\left<\phi\right>\left<\kappa_i\right>}.
\end{equation}

All simulation results presented in this work were generated using the Multiphase Flow with Interface eXchanges (MFIX) software that relies on a variable time step, staggered grid, finite-volume method for the solution of the two-fluid model equations~\citep{Syamlal1998}.  The iterative solution to the two-fluid model equations is obtained using the Semi-Implicit Method with Pressure Linked Equations, or the SIMPLE algorithm.  Due to the strong coupling between solid and fluid phases that arises due to the fluid-particle drag force, the Partial Elimination Algorithm of Spalding was used to effectively decouple the solution of the solid and fluid phase balance equations~\citep{Spalding1980}.  In addition, a second order Superbee discretization is employed for the convective terms that are present in the conservation equations for continuity, momentum and granular energy transport to limit the effects of numerical diffusion.     

\section{Preliminary Study}
\label{sec:prelim_study}
\begin{figure}
\includegraphics[width=5 in]{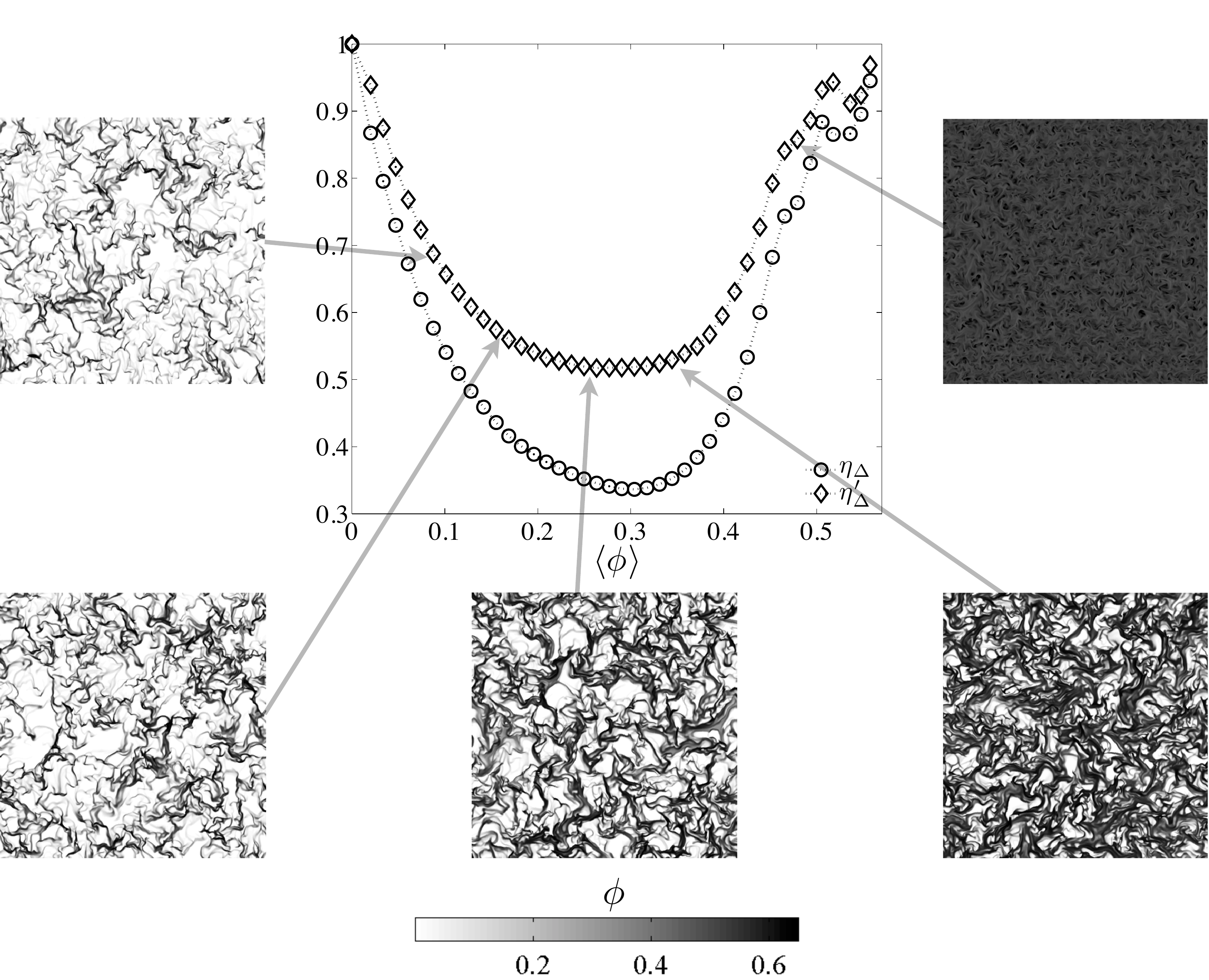}

\caption{A characteristic plot of the $\eta_{\Delta}'$ versus $\left<\phi\right>$ is shown with snapshots of the particle volume fraction field different domain-averaged volume fractions superimposed.  Here, $\hat{\Phi}=0.23$, $\hat{Sc}=1.00$, and $\hat{\Delta}=3.14$.}
\label{fig:eta_phi_dep}
\end{figure}

To illustrate the behavior of the \emph{cluster-scale} effectiveness factor as a function of $\left<\phi\right>$, simulation results are presented that were obtained using grid sizes that are sixteen times as large as the particle diameter.  This grid resolution was chosen to follow the simulations employed in the study of~\cite{Igci2008}.  \emph{Filtered} variables were obtained from these fine-grid simulations by moving filters of different sizes throughout the periodic domain.  Due to the statistical homogeneity of periodic domains we are able to collect thousands of samples and average them regardless of spatial position.  Moreover, by \emph{filtering} these fine-grid simulation results for various domain-averaged volume fractions the filter-averaged volume fraction dependence of the \emph{cluster-scale} effectiveness factor can be ascertained.  This \emph{filtering} procedure follows that outlined in the work of~\cite{Igci2008}.  The characteristic $\left<\phi\right>$ dependence of $\eta_{\Delta}$ and $\eta_{\Delta}'$ is presented in Figure~\ref{fig:eta_phi_dep}.  It is clear from Figure~\ref{fig:eta_phi_dep} that both $\eta_{\Delta}$ and $\eta_{\Delta}'$ are strong functions of $\left<\phi\right>$ retaining an inverted bell shape that approaches unity in the limit of small and large particle volume fraction.  However, there are noticeable quantitative differences between $\eta_{\Delta}$ and $\eta_{\Delta}'$ in Figure~\ref{fig:eta_phi_dep} with the maximum deviation being near the minimum in both curves.  The minimum value of $\eta_{\Delta}\approx 0.33$, while the minimum value of $\eta_{\Delta}'\approx 0.53$.  Therefore, at this resolution, the non-local correction to the cluster-scale effectiveness factor contributes as much as $37 \%$ to the value of the \emph{cluster-scale} effectiveness factor near the minimum in the curves given in Figure~\ref{fig:eta_phi_dep}.  While the quantitative differences between $\eta_{\Delta}$ and $\eta_{\Delta}'$ are functions of $\hat{\Delta}$ and other model parameters, this example provides a scale of the difference between the two effectiveness factors.  In this work we have decided to model $\eta_{\Delta}'$ because it is directly related the small scale fluctuations in $\left<\phi\right>$ and $\left<\chi_{g_{i}}\right>$, while $\eta_{\Delta}$ is dependent on non-local variations in $\left<\phi\right>$ and $\left<\chi_{g_{i}}\right>$ (see Appendix).  Physical intuition for the decrease in the \emph{cluster-scale} effectiveness factor from unity can be obtained by observing the characteristic clustering patterns that are observed at different domain-averaged volume fractions, which are superimposed in Figure~\ref{fig:eta_phi_dep}.  At low volume fractions there are a few small isolated clusters throughout the periodic domain, and as such, the cluster-scale effectiveness factor will begin to deviate from unity.  As the volume fraction increases the frequency of these clusters increases thus making the effective contacting between gas and particle phases poor.  Near the minimum value in the $\eta_{\Delta}'$ curve, clusters begin to span the periodic domain, and the solid phase changes from the dispersed to the continuous phase.  As volume fraction continues to increase the gas-particle flow becomes more homogeneous, and due to this homogeneity the cluster-scale effectiveness factor increases toward unity in the limit of high particle volume fraction.

\begin{figure}

\subfigure[]{\includegraphics[width=2.5 in]{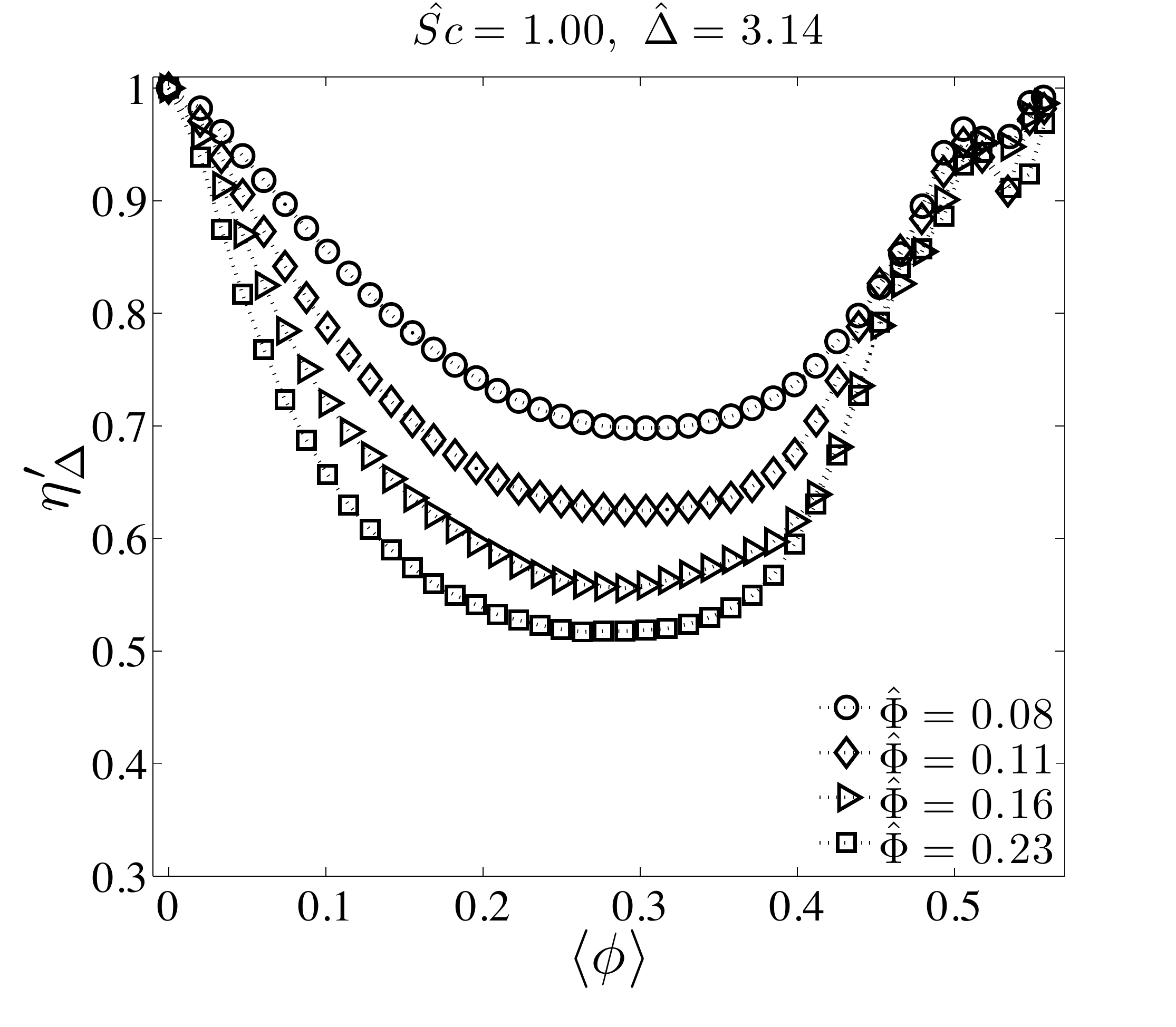} }
\subfigure[]{\includegraphics[width=2.5 in]{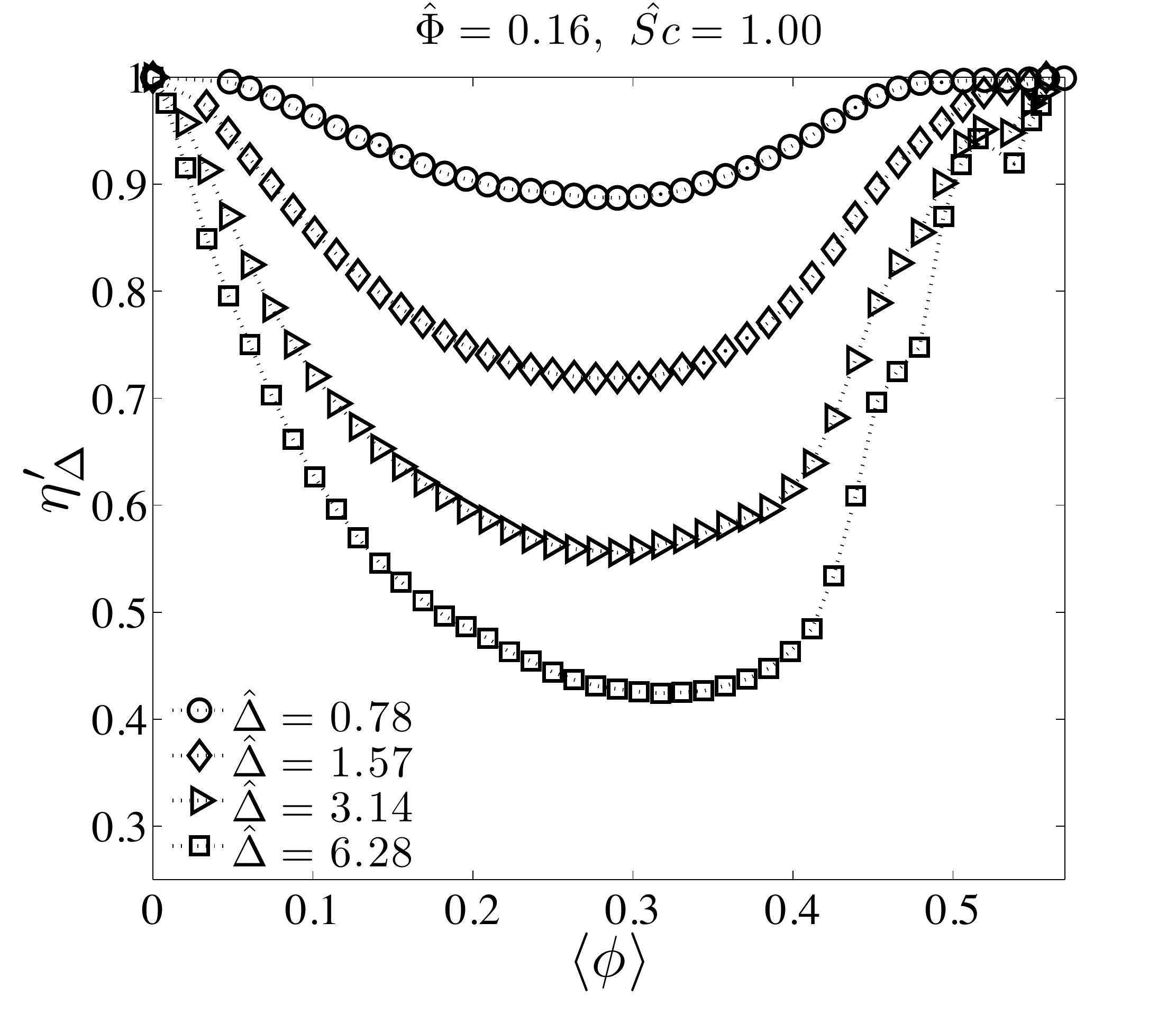} }
\begin{center}
\subfigure[]{\includegraphics[width=2.5 in]{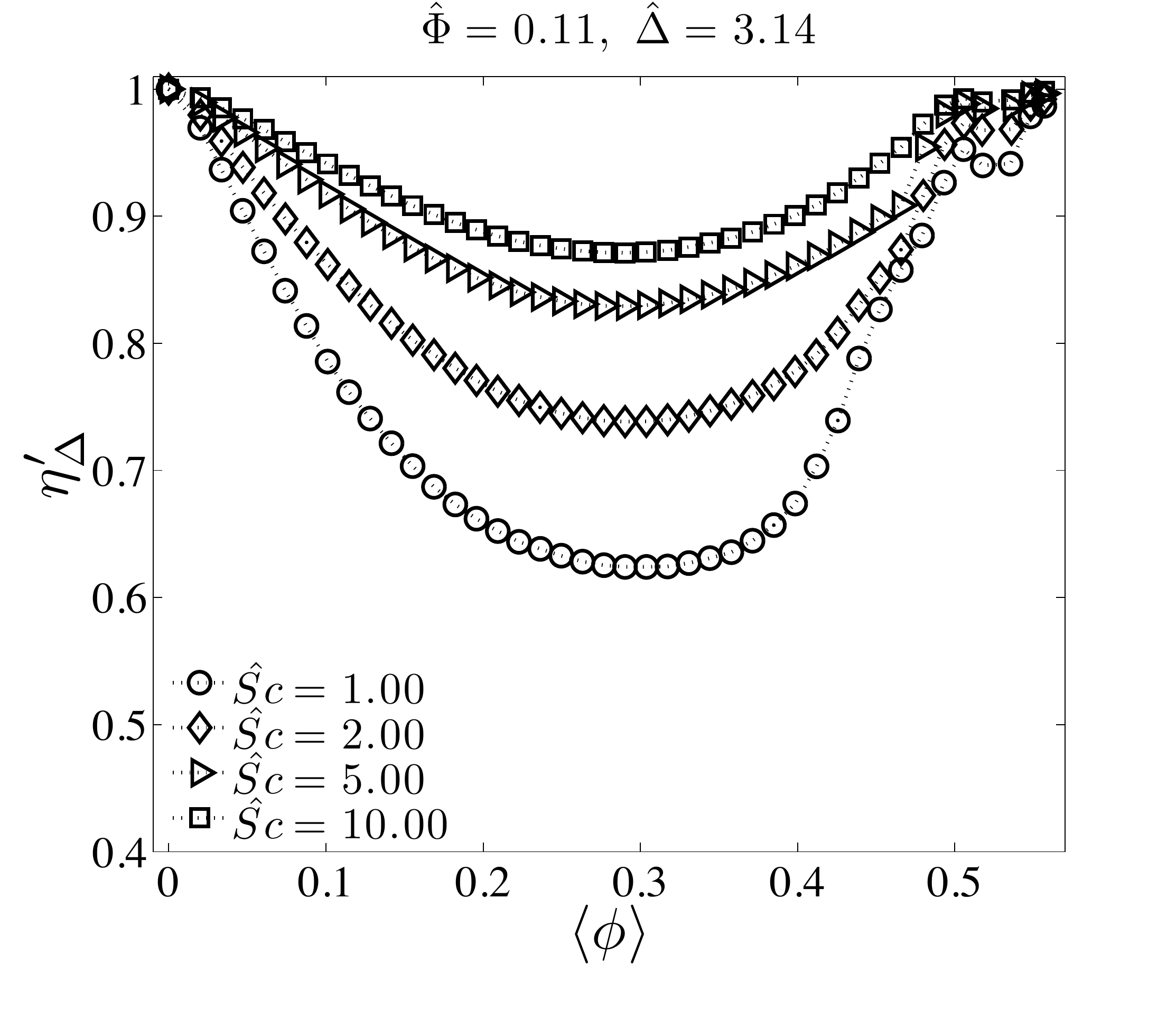} }
\end{center}
\caption{ The \emph{cluster} effectiveness factor $\eta_{\Delta}'$ as a function of filtered particle volume fraction $\left<\phi\right>$.  (a) $\eta_{\Delta}'$ versus $\left<\phi\right>$ for different values of the meso-scale Thiele modulus $\hat{\Phi}$ for fixed values of the meso-scale Schmidt number $\hat{Sc}$ and dimensionless filter size $\hat{\Delta}$; (b) $\eta_{\Delta}'$ versus $\left<\phi\right>$ for different values of $\hat{\Delta}$ for fixed values of $\hat{Sc}$ and $\hat{\Phi}$; and (c) $\eta_{\Delta}'$ versus $\left<\phi\right>$ for different values of $\hat{Sc}$ for fixed values of $\hat{\Delta}$ and $\hat{\Phi}$.}
\label{fig:eta_curves}
\end{figure}

In Figure~\ref{fig:eta_curves} (a) the \emph{cluster-scale} effectiveness factor is shown as a function of $\left<\phi\right>$ for four different $\hat{\Phi}$ values at fixed values of the $\hat{Sc}$ and $\hat{\Delta}$.  The depth of the inverted bell shape curve in $\left<\phi\right>$ is an increasing function of $\hat{\Phi}$.  The values of $\hat{\Phi}$ given in Figure~\ref{fig:eta_curves} (a) are substantially smaller than unity, but there is a marked change in $\eta_{\Delta}'$ from unity.  This may seem peculiar, but the small values of $\hat{\Phi}$ arise as a result of the fact that the length scale used to determine $\hat{\Phi}$ is the particle diameter $d$.  A more fitting length scale is that associated with a cluster.  However, the size of a cluster emerges as a result of fine-grid two-fluid model simulations, and cannot be considered as an input parameter. It is for this reason that we choose the particle diameter as the relevant length scale for $\hat{\Phi}$ rather than the length scale of a cluster.  In Figure~\ref{fig:eta_curves} (b) the $\left<\phi\right>$ dependence of the \emph{cluster-scale} effectiveness factor is given for four different values of the dimensionless filter size $\hat{\Delta}$ at fixed $\hat{\Phi}$ and $\hat{Sc}$.  The departure of the effectiveness factor from unity is also an increasing function of $\hat{\Delta}$.  However, Figure~\ref{fig:eta_curves} (c) shows that the change in the \emph{cluster-scale} effectiveness factor from unity is a decreasing function of $\hat{Sc}$ when keeping $\hat{\Phi}$ and $\hat{\Delta}$ constant.  While one might expect that increasing the $\hat{Sc}$ value should increase the departure in $\eta_{\Delta}'$ from unity, we observe a decrease in this departure due to the fact that we are demanding $\hat{\Phi}$ to remain constant in Figure~\ref{fig:eta_curves} (c).  In these simulations the variation of $\hat{Sc}$ is achieved by varying the value of $D$.  In order to maintain a constant $\hat{\Phi}$ the effective rate constant $k_\text{eff}$ must be changed.  Therefore, varying $\hat{Sc}$ at constant $\hat{\Phi}$ requires the variation in the effective rate constant.  It is this coupled variation that brings about the somewhat surprising the dependence of $\eta_{\Delta}'$ on $\hat{Sc}$.  

\section{Grid Resolution Effect}

\begin{figure}
\subfigure[]{\includegraphics[width=2.5 in]{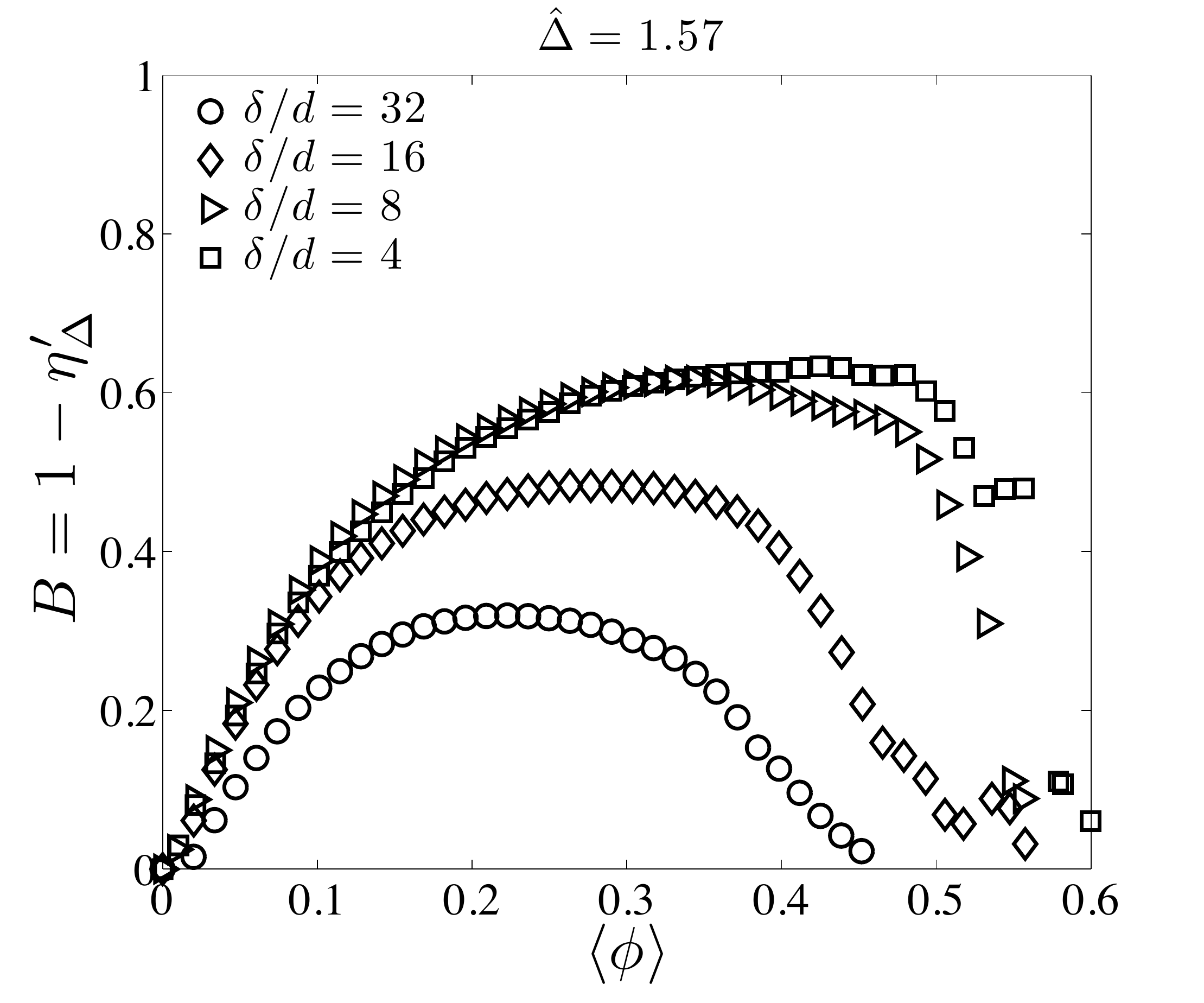} }
\subfigure[]{\includegraphics[width=2.5 in]{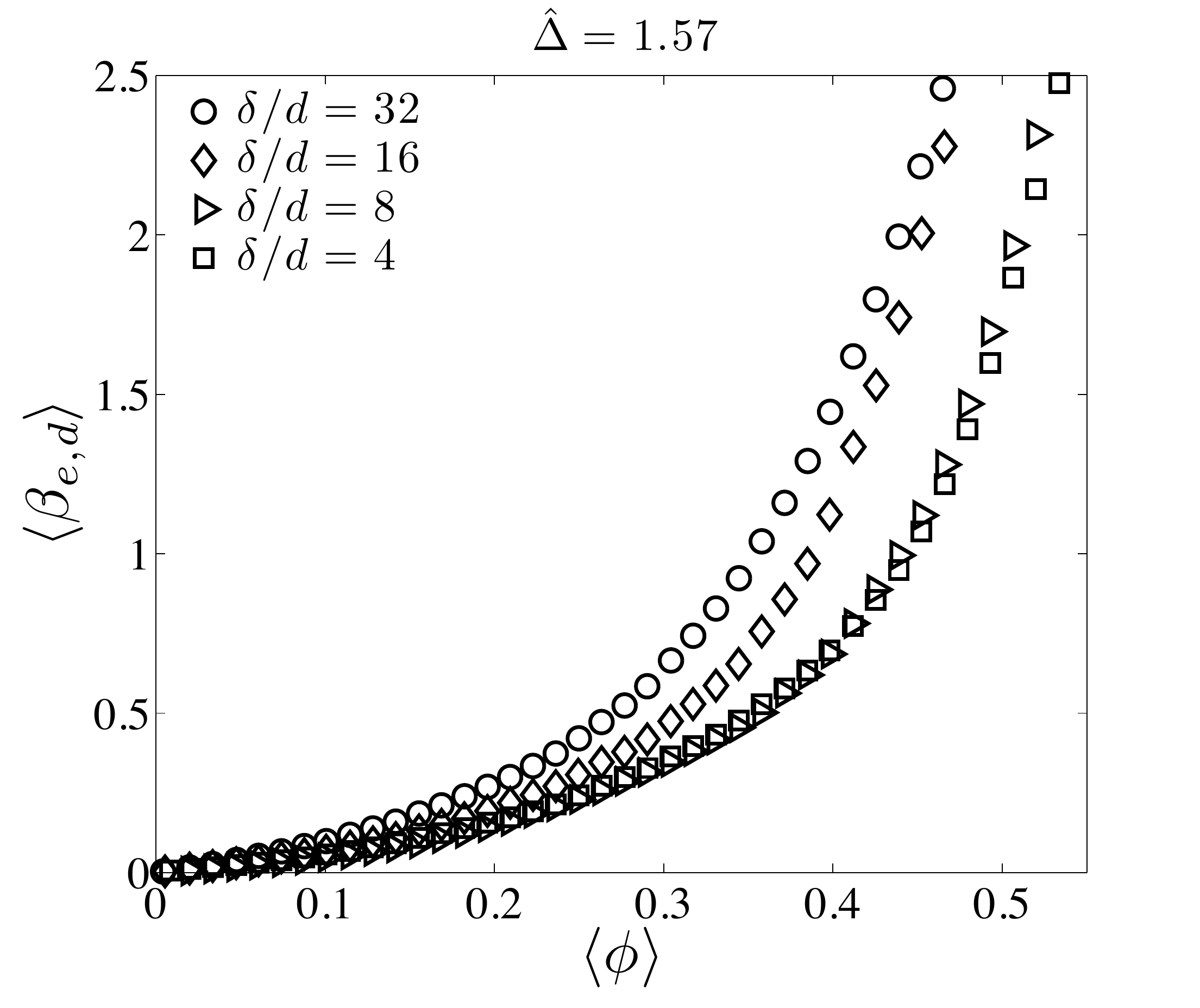} }
\caption{Grid size $\delta$ dependence of filtered quantities. (a) $B=1-\eta_{\Delta}'$ is plotted versus $\left<\phi\right>$ for several different grid resolutions.  (b) The dimensionless \emph{filtered} drag coefficient $\left<\beta_{e,d}\right>$ is plotted versus $\left<\phi\right>$ for several different grid resolutions to show the degree to which hydrodynamic variables are fully resolved.  Here, $\hat{\Phi}=0.23$ and $\hat{Sc}=1.00$.}
\label{fig:eta_grid_dep}
\end{figure}

When deducing a \emph{filtered} model from a fine-grid simulation, it is necessary to ensure that \emph{filtered} statistics are independent of grid size $\delta$.  To that end, the dependence of $B=1-\eta_{\Delta}'$ on $\left<\phi\right>$ is presented for four different grid resolutions in Figure~\ref{fig:eta_grid_dep} (a) at fixed $\hat{\Phi}$, $\hat{\Delta}$, and $\hat{Sc}$.  The dependence of $\eta_{\Delta}'$ on grid size $\delta$ is substantial, and only begins to saturate when the grid size is around $4-8$ particle diameters.  Figure 2 (b) shows the variation of the dimensionless \emph{filtered} drag coefficient $\left<\beta_{e,d}\right>$ as a function of $\left<\phi\right>$ for four different grid resolutions.  Here the dimensionless \emph{filtered} drag coefficient is defined consistent with the earlier work of~\cite{Igci2008}
\begin{equation}
\left<\beta_{e,d}\right>=\frac{v_t}{\rho_s |\boldsymbol{g}|}\left(\frac{\left<f_{D_{y}}\right>-\left<\phi 'dp_g'/dy\right>}{u_y-v_y}\right)
\label{eq:beta_def} 
\end{equation}
where the subscript $y$ indicates that the drag coefficient is inferred from the drag force and pressure fluctuation terms directed parallel to gravity.  At grid sizes of sixteen particle diameters $\left<\beta_{e,d}\right>$ seems to exhibit grid independence for $\left<\phi\right> < 0.30$, with some grid dependence emerging at higher volume fractions.  Figures 3 (a) and (b) illustrate the sensitive grid dependence of the \emph{cluster-scale} effectiveness factor when compared to other \emph{filtered} quantities like the fluid-particle drag coefficient.  Consequently, even finer grid resolutions are required when simulating reacting gas-particle flow when compared to non-reacting systems.  This observation further supports the need for the development of \emph{filtered} models for accurate coarse-grid simulation of reacting systems.

\begin{figure}
\subfigure[]{\includegraphics[width=2.5 in]{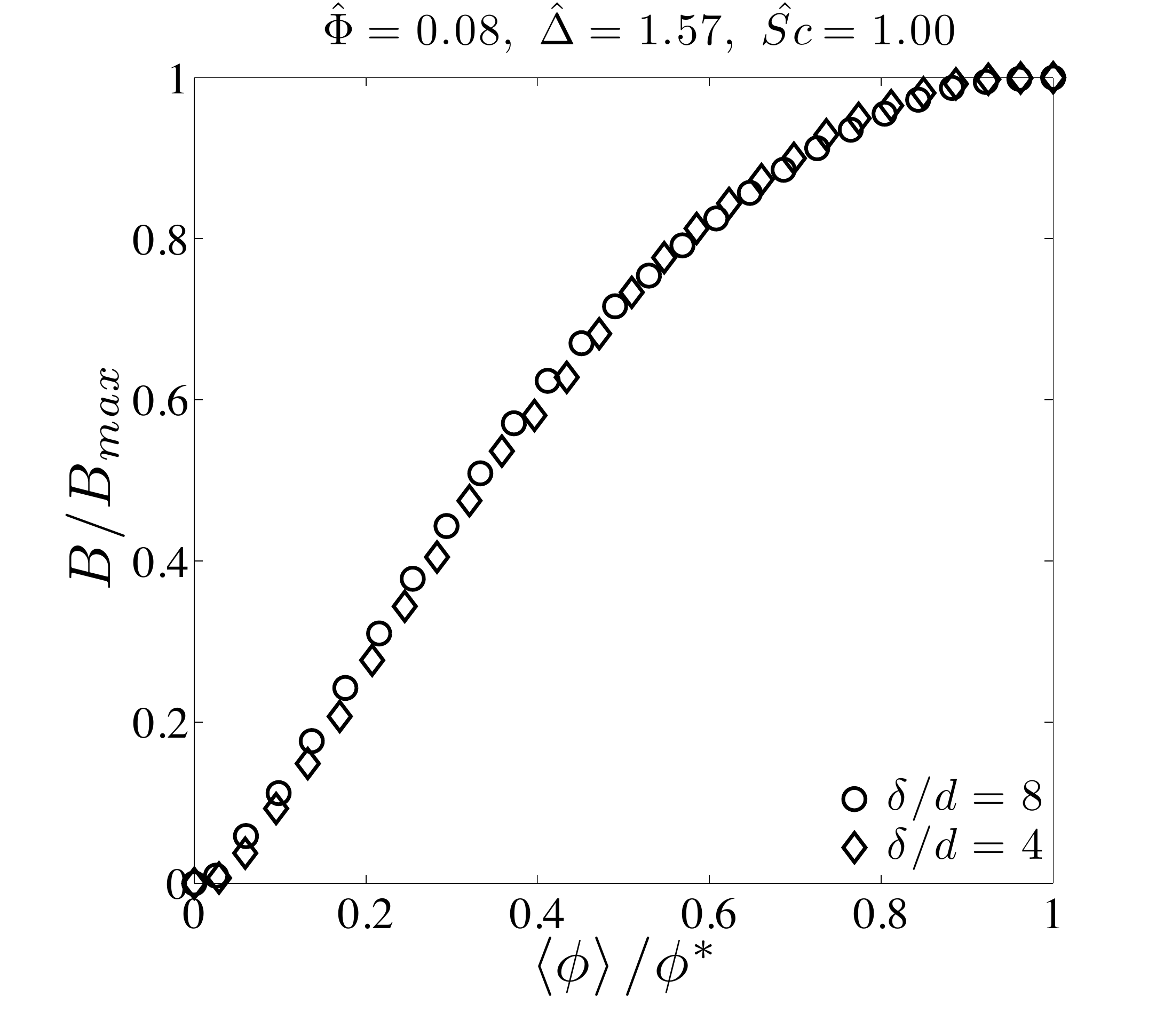} }
\subfigure[]{\includegraphics[width=2.5 in]{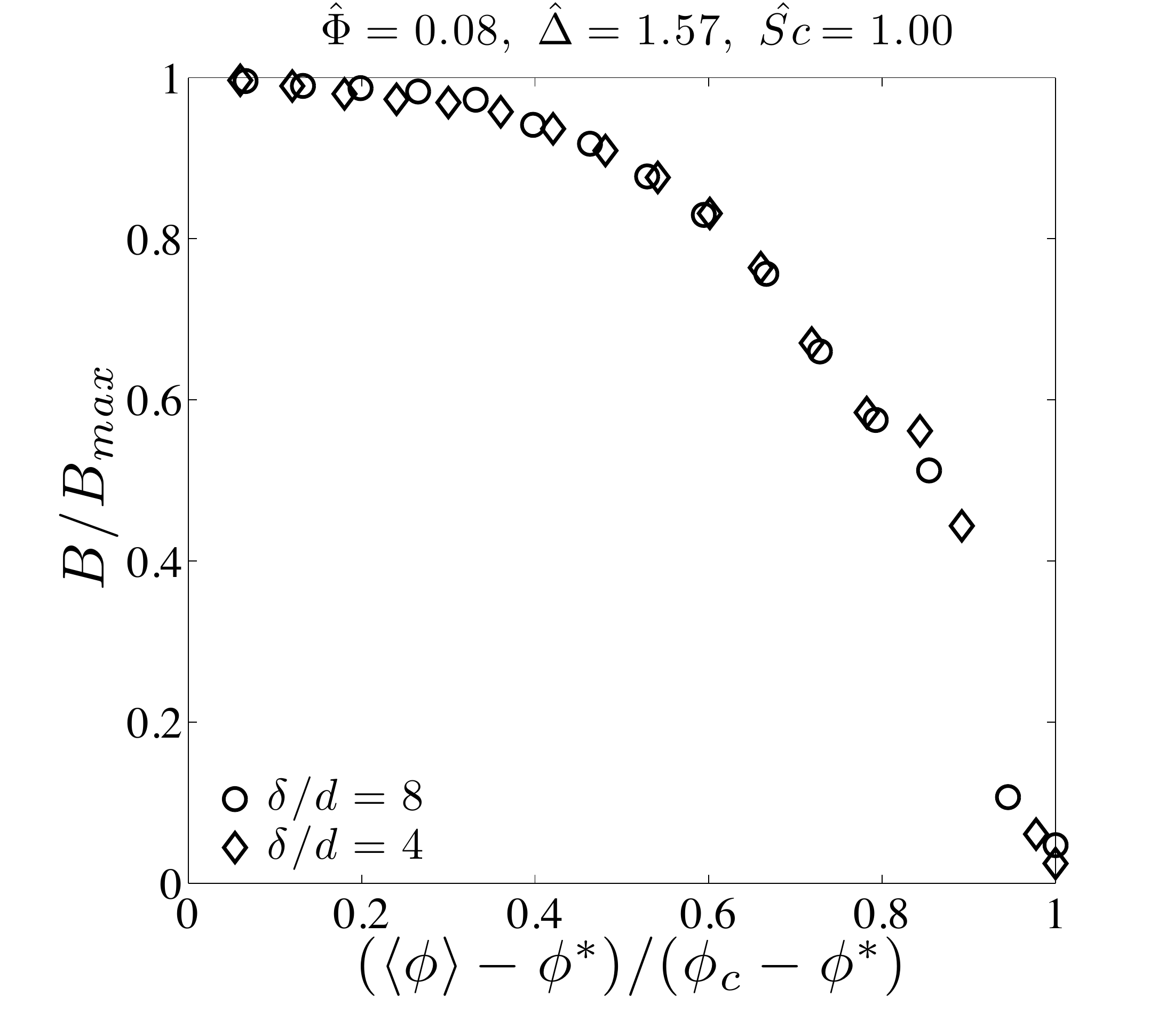} }
\subfigure[]{\includegraphics[width=2.5 in]{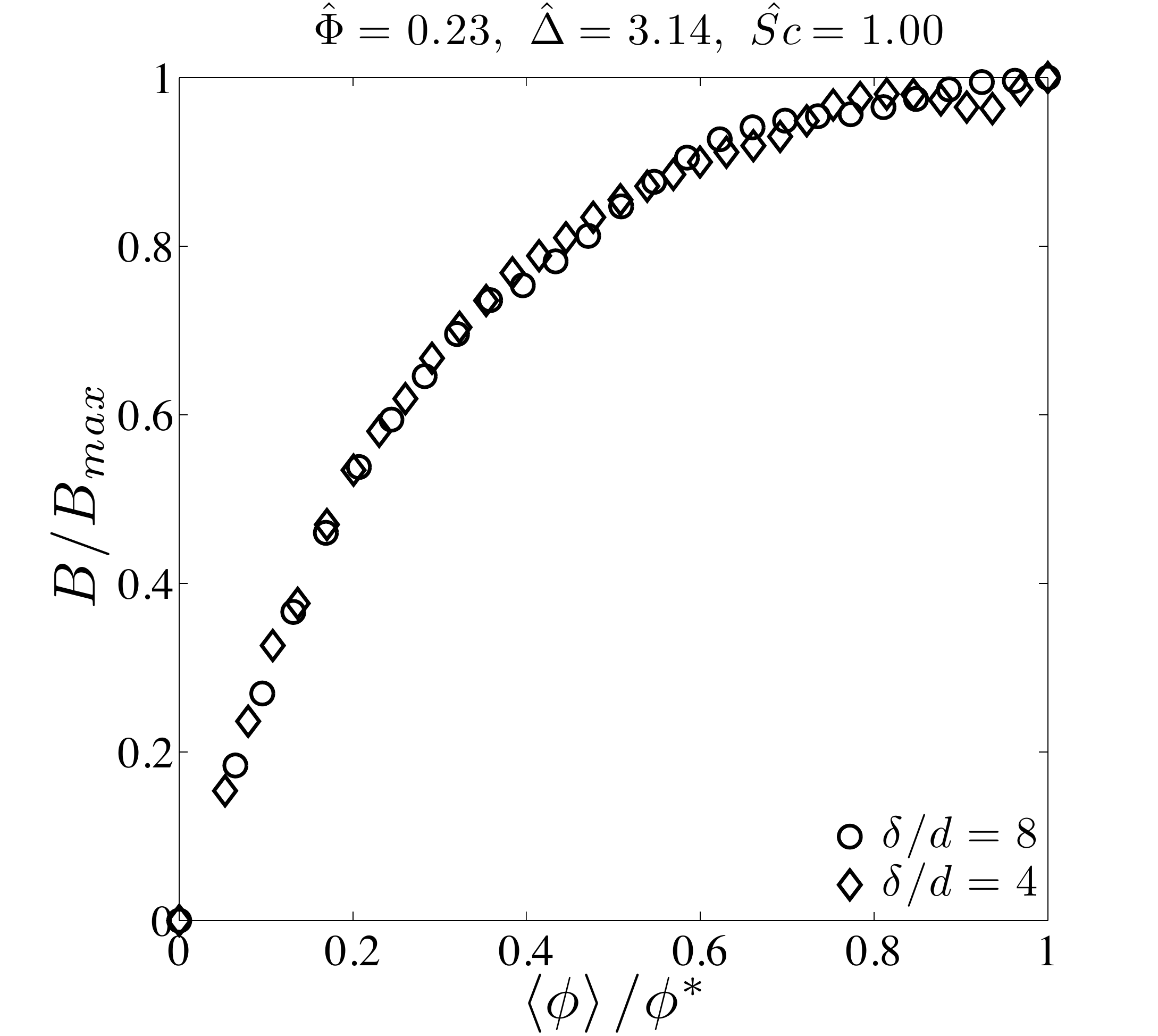} }
\subfigure[]{\includegraphics[width=2.5 in]{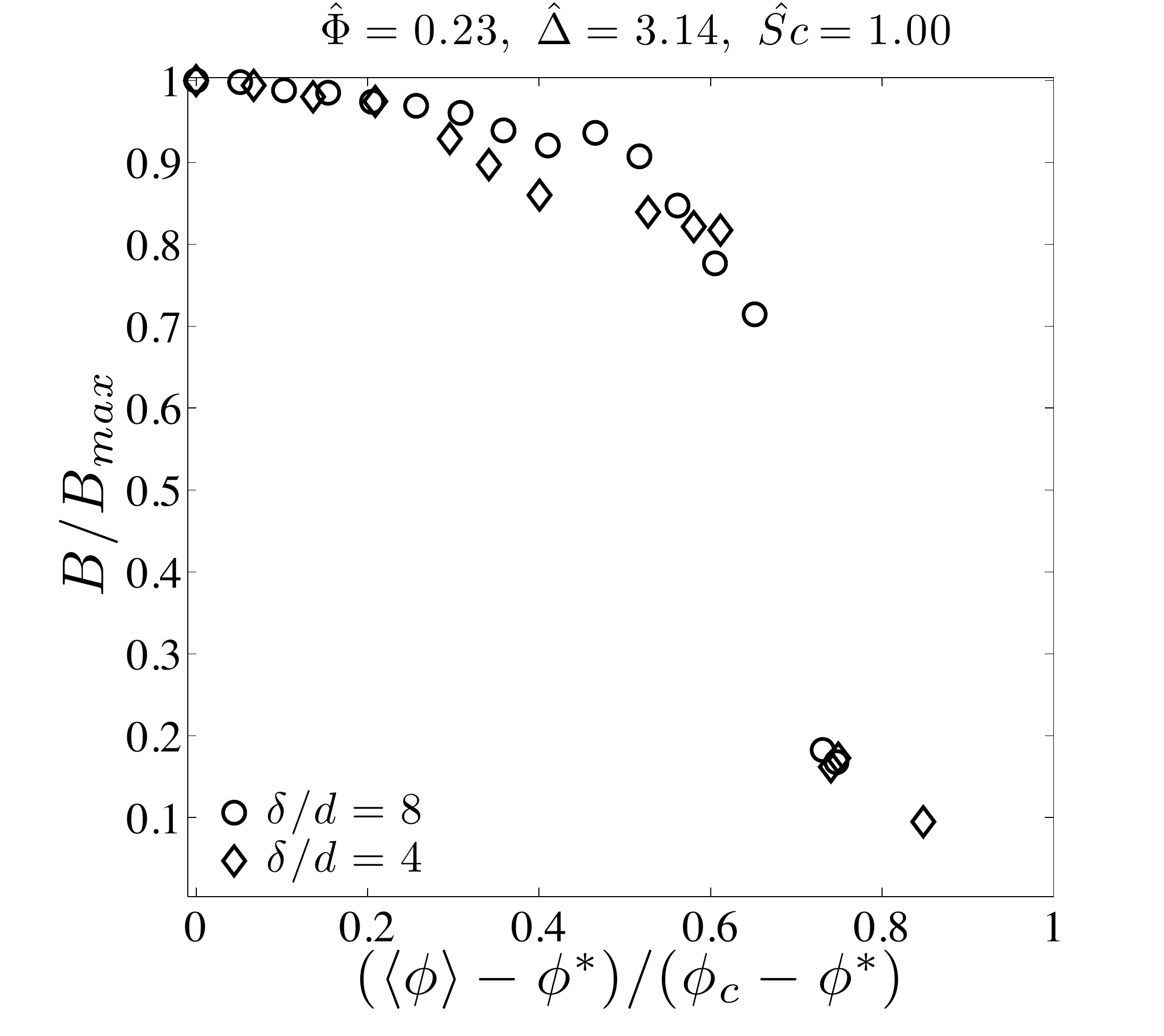} }
\caption{Here we demonstrate the grid size independence of the \emph{cluster-scale} effectiveness factor curves when appropriately scaled, and the volume fraction dependence is considered separately to left and right of $\phi^*$.  In (a) and (c) we plot $B=1-\eta_{\Delta}'$ scaled by its maximum value $B_{max}$ as a function of $\left<\phi\right>/\phi^*$ for the finest two grid resolutions in this study.  In (b) and (d) we plot $B/B_{max}$ as a function of $\left(\left<\phi\right>-\phi^*\right)/\left(\phi_c-\phi^*\right)$. }
\label{fig:rescaled_plots}
\end{figure}

Due to the sensitivity of $\eta_{\Delta}'$ with respect to grid size, we seek to develop a model for the \emph{cluster-scale} effectiveness factor extrapolated to the limit of infinite grid resolution.  In Figures~\ref{fig:rescaled_plots} (a) and (c), plots of $B/B_{max}$ versus $\left<\phi\right>/\phi^*$ are given for the finest two grid resolutions presented in this study.  Here, $B_{max}$ is the maximum value of $B$, and $\phi^*$ is the \emph{filtered} volume fraction at which the maximum value of $B$ occurs.  From inspection of Figures~\ref{fig:rescaled_plots} (a) and (c) it is clear that the grid dependence of the volume fraction variation in $B$ can be removed by rescaling $B$ by $B_{max}$ and $\left<\phi\right>$ by $\phi^*$ for all $\left<\phi\right> < \phi^*$.   In addition, it is demonstrated in Figures~\ref{fig:rescaled_plots} (b) and (d) that the grid dependence of the volume fraction variation in $B$ can be removed for all $\left<\phi\right> >\phi^*$ by plotting $B/B_{max}$ versus $\left(\left<\phi\right>-\phi^*\right)/\left(\phi_c-\phi^*\right)$.  Here $\phi_c$ is the volume fraction at which the value of $B$ reaches zero. Due to the grid-independent behavior observed in Figures~\ref{fig:rescaled_plots} (a)$-$(d)  a filtered model utilizing a piecewise description of the variation in $B/B_{max}$ with particle volume fraction about $\phi^*$ is developed, while extrapolating the values of $B_{max}$, $\phi^*$, and $\phi_c$ to infinite resolution.

\begin{figure}
\subfigure[]{\includegraphics[width=2.5 in]{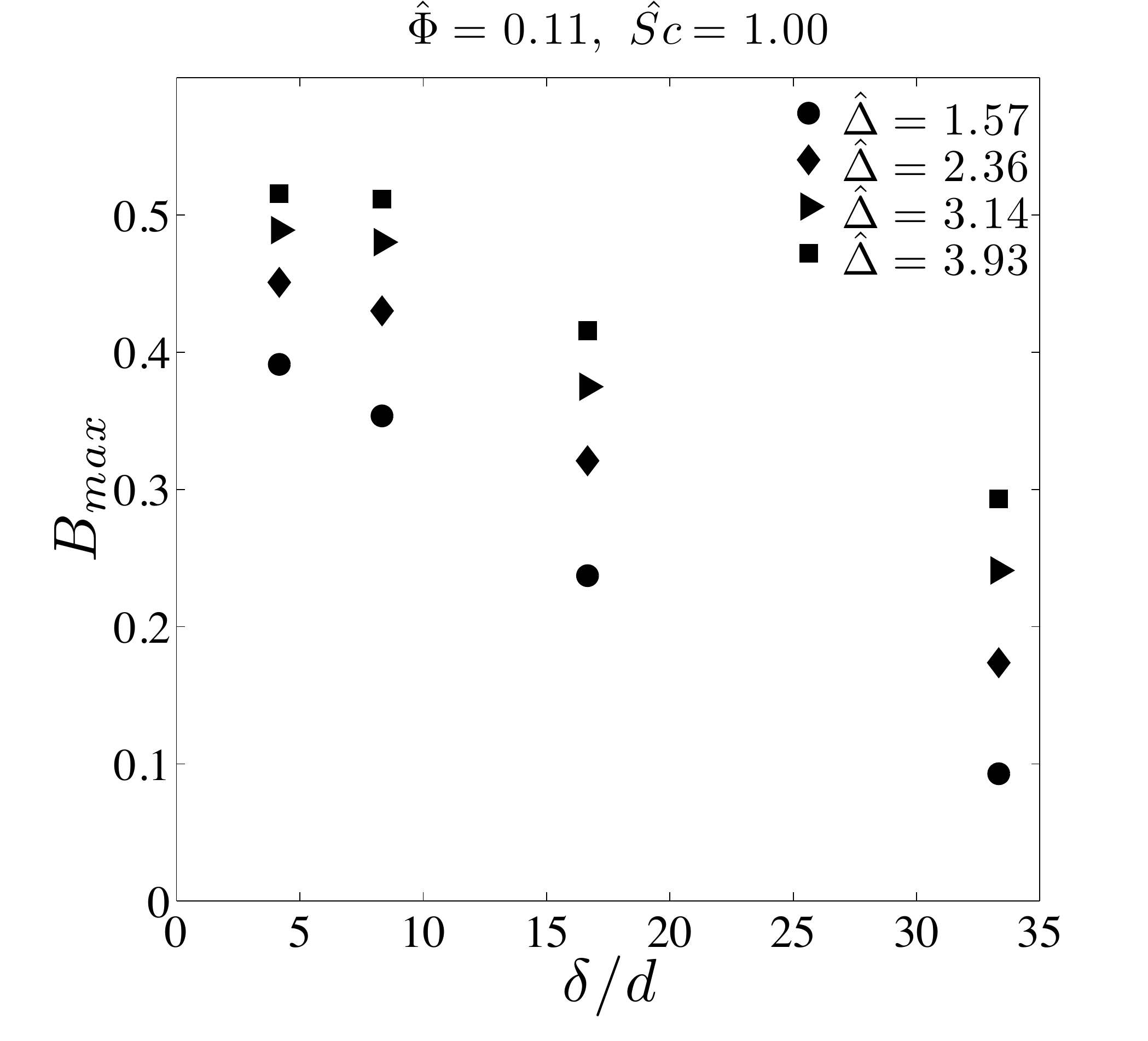} }
\subfigure[]{\includegraphics[width=2.5 in]{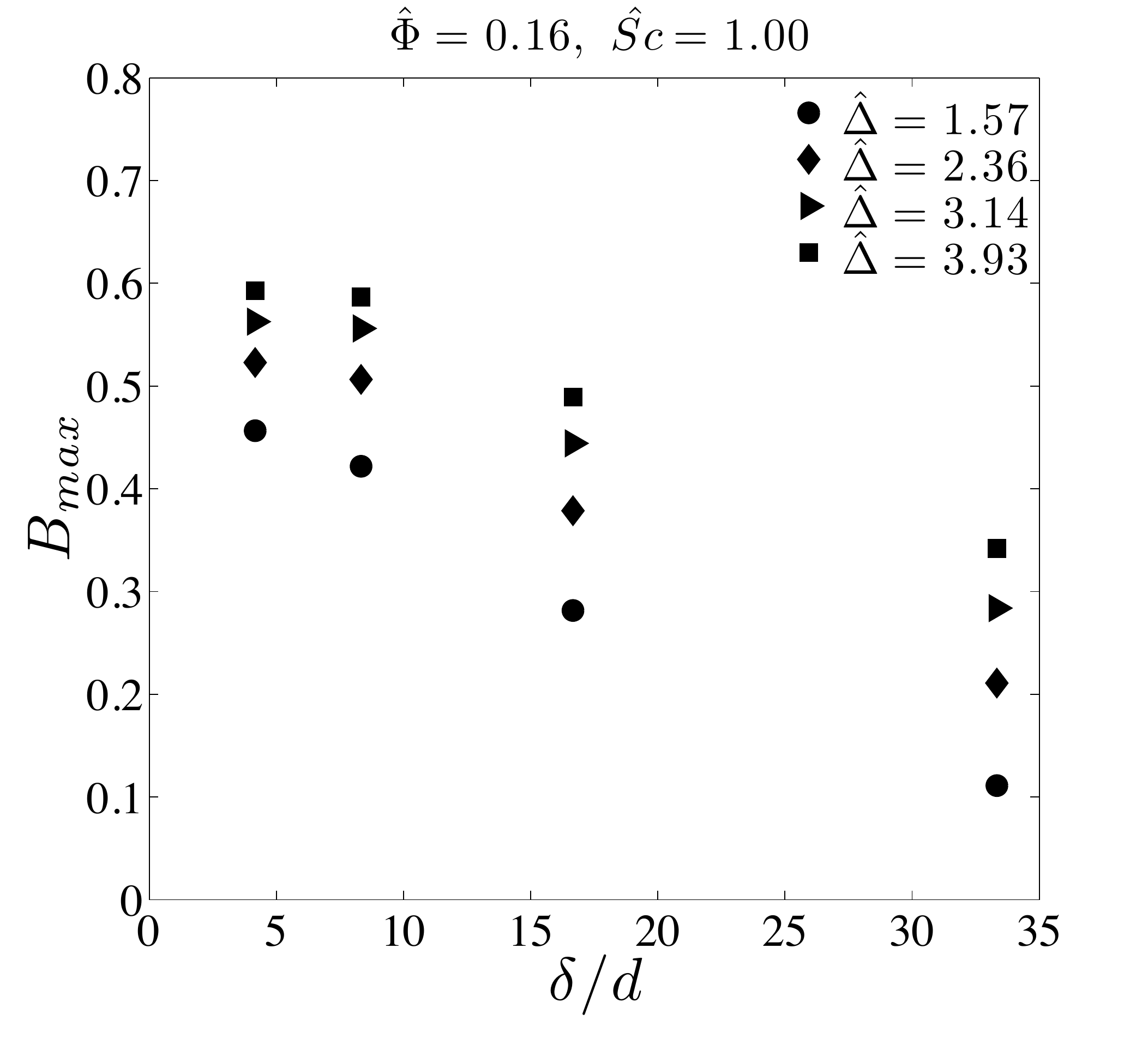} }
\caption{We present $B_{max}$ plotted versus $\delta/d$ for several different values of $\hat{\Delta}$.  (a) $\hat{\Phi}=0.11$ and $\hat{Sc}=1.00$; (b) $\hat{\Phi}=0.16$ and $\hat{Sc}=1.00$.}
\label{fig:extrap_B}
\end{figure}

In Figure~\ref{fig:extrap_B} (a) and (b) the grid size dependence of $B_{max}$ is presented for two different values of $\hat{\Phi}$ and a variety of different filter sizes.  The values of $B_{max}$ are clearly saturating as the value of $\delta$ approaches zero.  In order to provide an accurate value of $B_{max}$ to use in our \emph{filtered} reaction rate model, the value of $B_{max}$ at infinite resolution is determined via a Richardson Extrapolation~\citep{Roache1998}.  The values of $\phi^*$ and $\phi_c$ were observed to vary linearly with grid resolution, and a linear extrapolation was performed to ascertain the infinitely resolved estimates of $\phi^*$ and $\phi_c$.

\section{Extrapolated results and \emph{filtered} model}

\begin{figure}
\subfigure[]{\includegraphics[width=2.5 in]{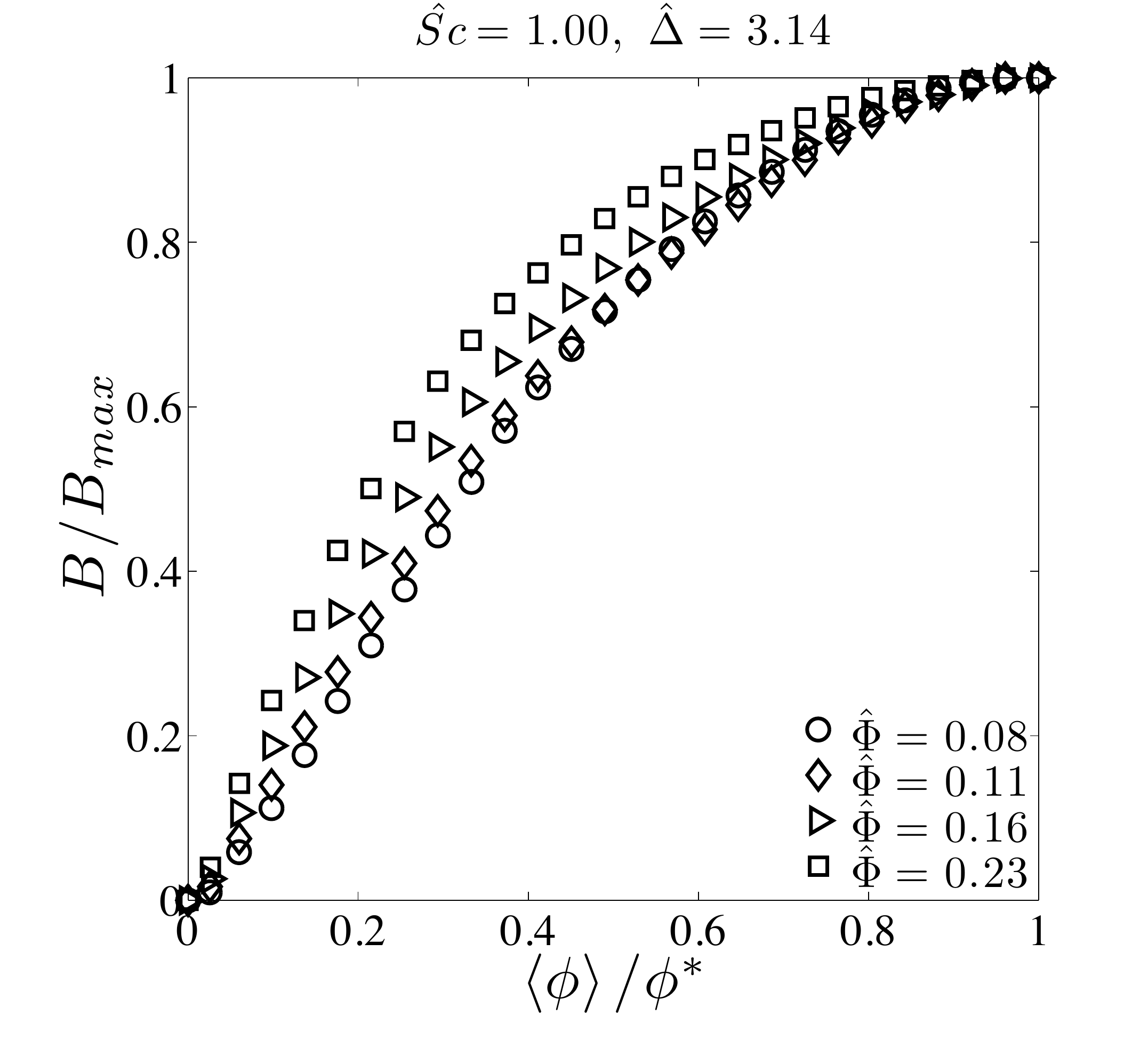} }
\subfigure[]{\includegraphics[width=2.5 in]{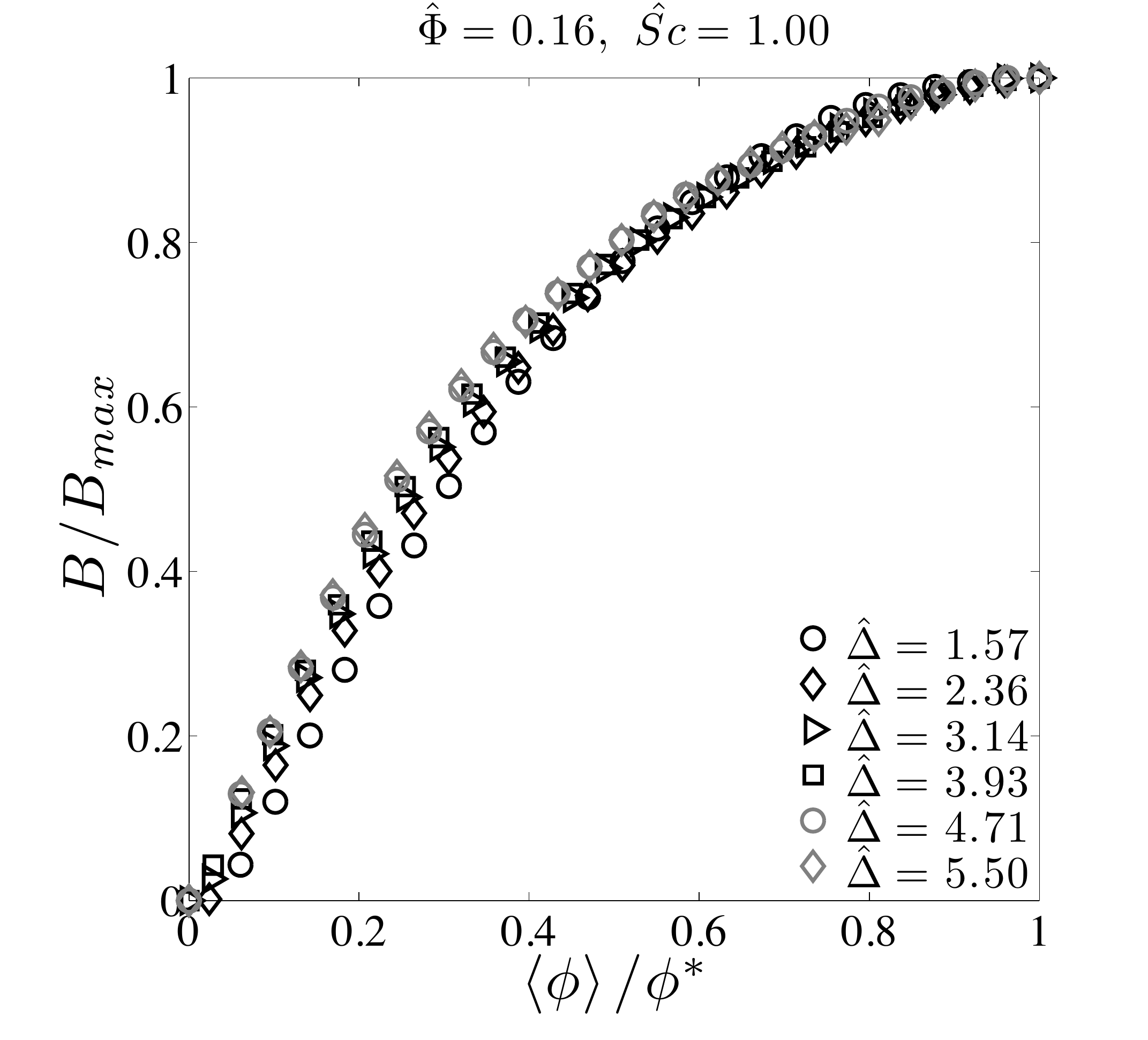} }
\subfigure[]{\includegraphics[width=2.5 in]{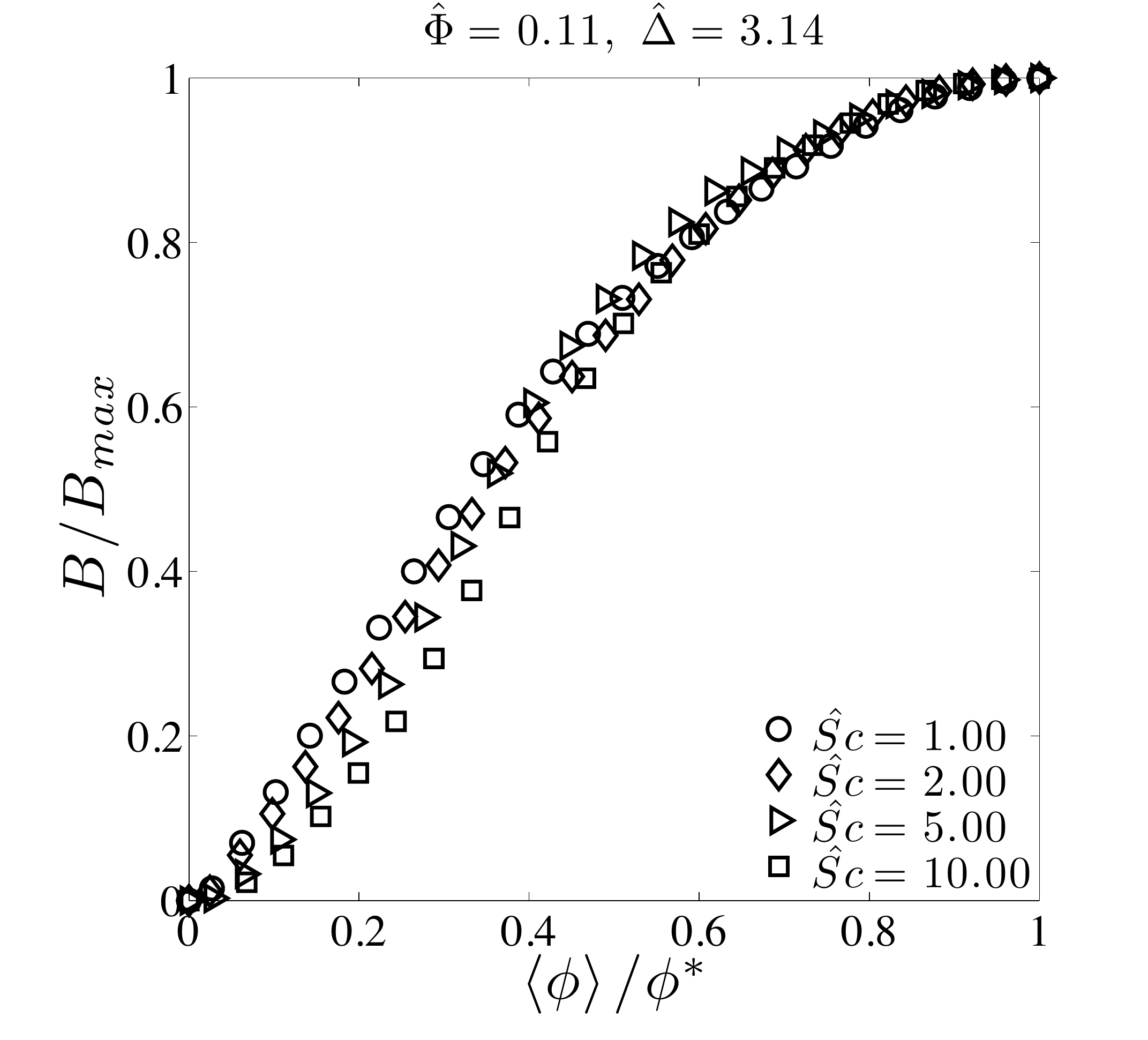} }
\subfigure[]{\includegraphics[width=2.5 in]{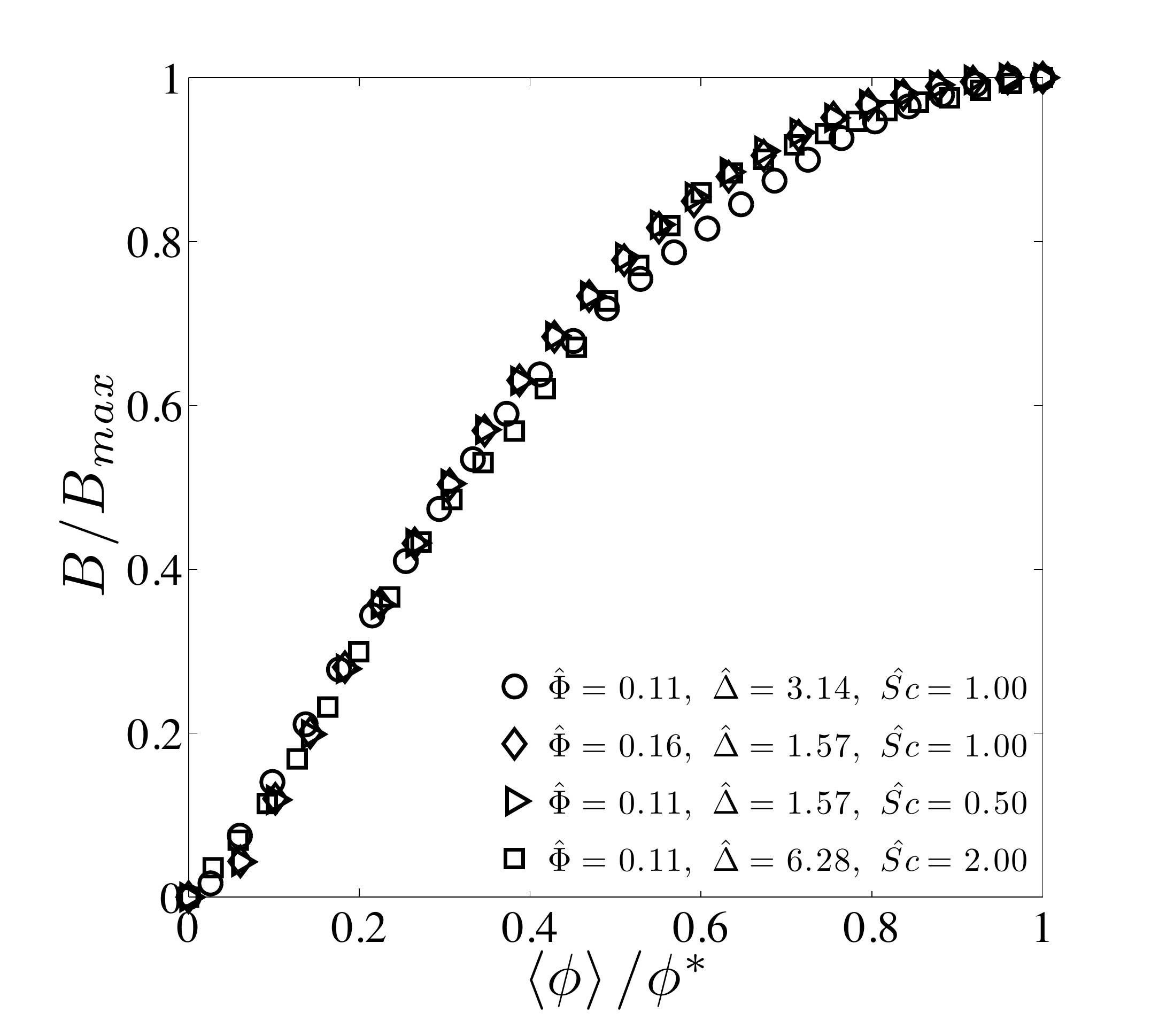} }
\caption{Dependence of $B/B_{max}$ on $\left<\phi\right>/\phi^*$ for various values (a) $\hat{\Phi}$, (b) $\hat{\Delta}$, and (c)$\hat{Sc}$.  In (d) we present $B/B_{max}$ versus $\left<\phi\right>/\phi^*$ where $\hat{\Phi}^2\hat{\Delta}/\hat{Sc}$ is constant.}
\label{fig:vol_fr_scaling_delta_left}
\end{figure}

In the previous section, grid independence of the \emph{cluster-scale} effectiveness factor was observed by plotting $B/B_{max}$ against scaled volume fraction coordinates that differ depending on whether $\left<\phi\right>$ is greater than or less than $\phi^*$.  Motivated by this observation, we seek to model the volume fraction variation of $B/B_{max}$ via a piecewise function about $\phi^*$.  In Figures~\ref{fig:vol_fr_scaling_delta_left} (a)$-$(c) plots of $B/B_{max}$ versus $\left<\phi\right>/\phi^*$ are presented for different values of $\hat{\Phi}$, $\hat{\Delta}$, and $\hat{Sc}$.  The shape of the curve of $B/B_{max}$ is clearly altered by changes in $\hat{\Phi}$ and $\hat{Sc}$, with only mild changes in the volume fraction dependence of $B/B_{max}$ at different grid values of $\hat{\Delta}$.  In Figure~\ref{fig:vol_fr_scaling_delta_left} (d) the variation in $B/B_{max}$ with $\left<\phi\right>/\phi^*$ is shown to collapse by plotting all data with the same value of $\hat{\Phi}^2\hat{\Delta}/\hat{Sc}$ together.

\begin{figure}
\subfigure[]{\includegraphics[width=2.5 in]{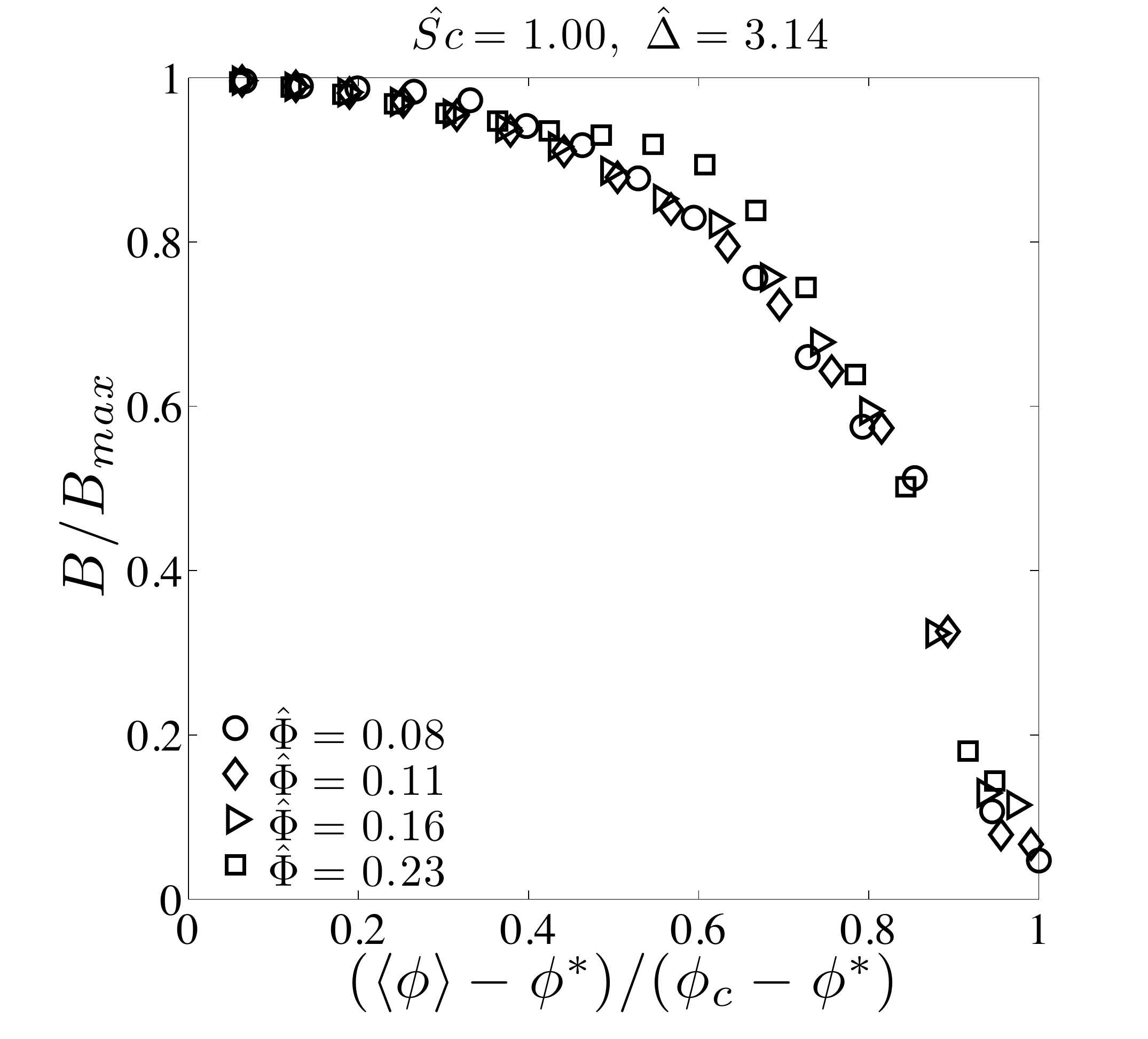} }
\subfigure[]{\includegraphics[width=2.5 in]{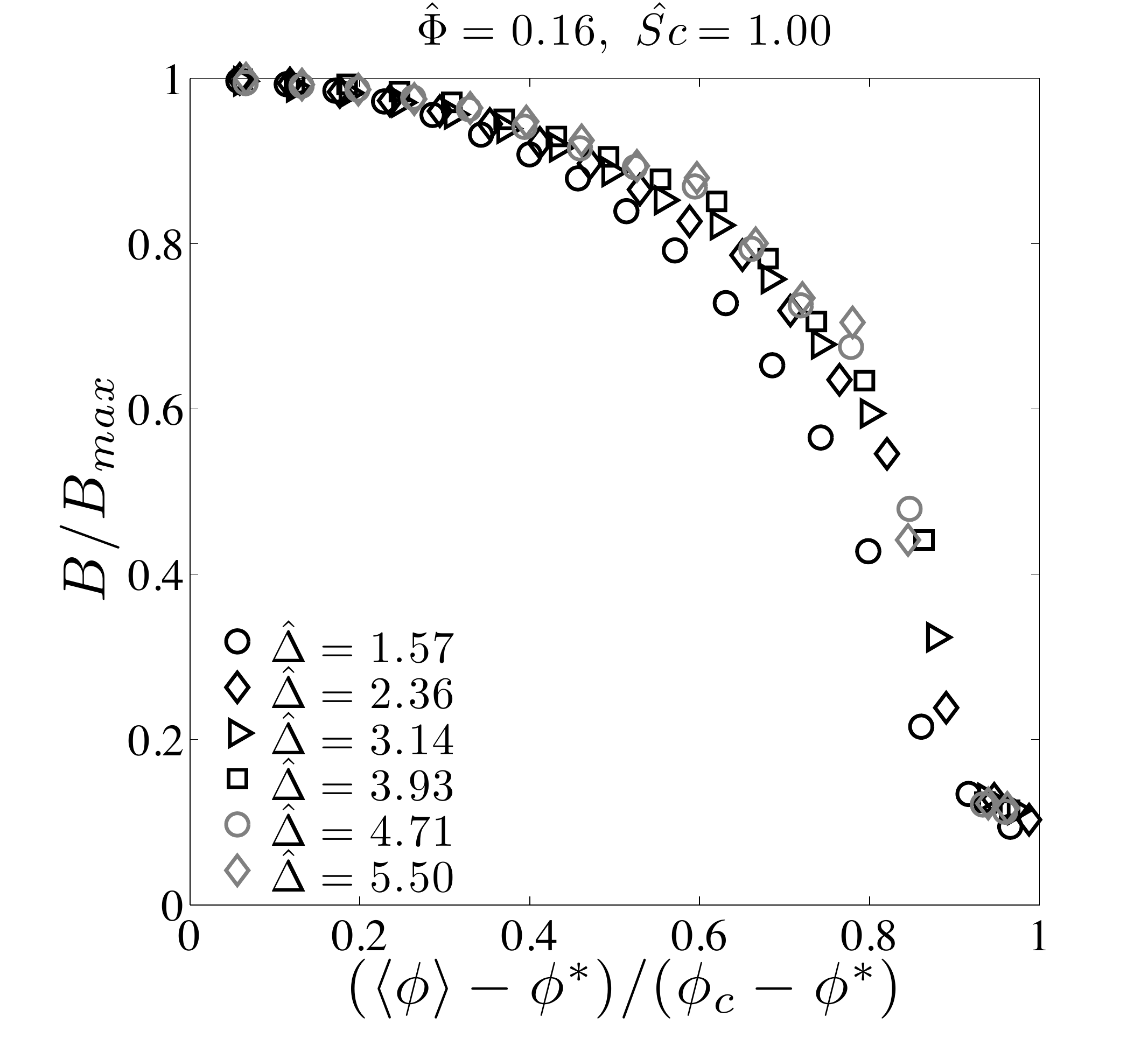} }
\subfigure[]{\includegraphics[width=2.5 in]{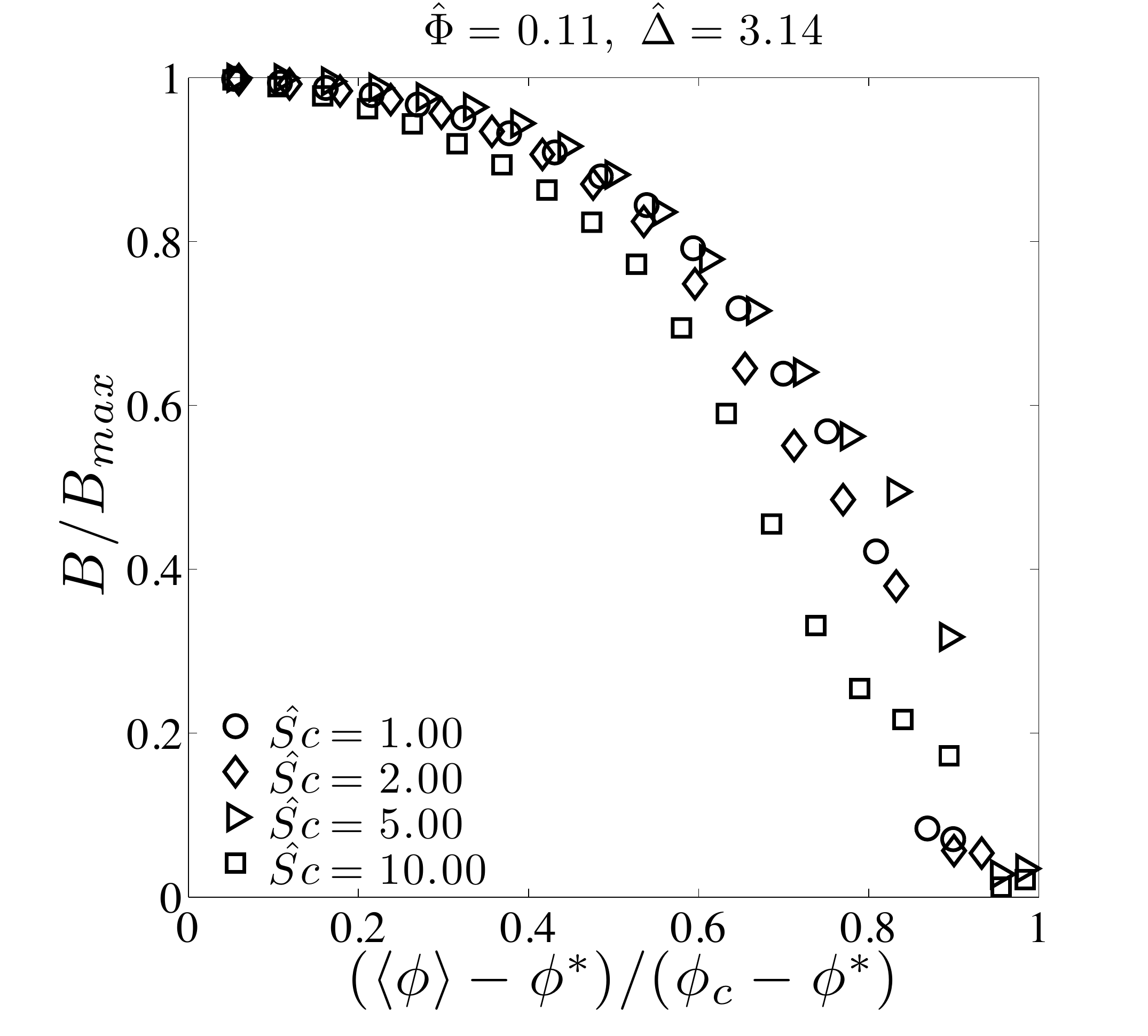} }
\subfigure[]{\includegraphics[width=2.5 in]{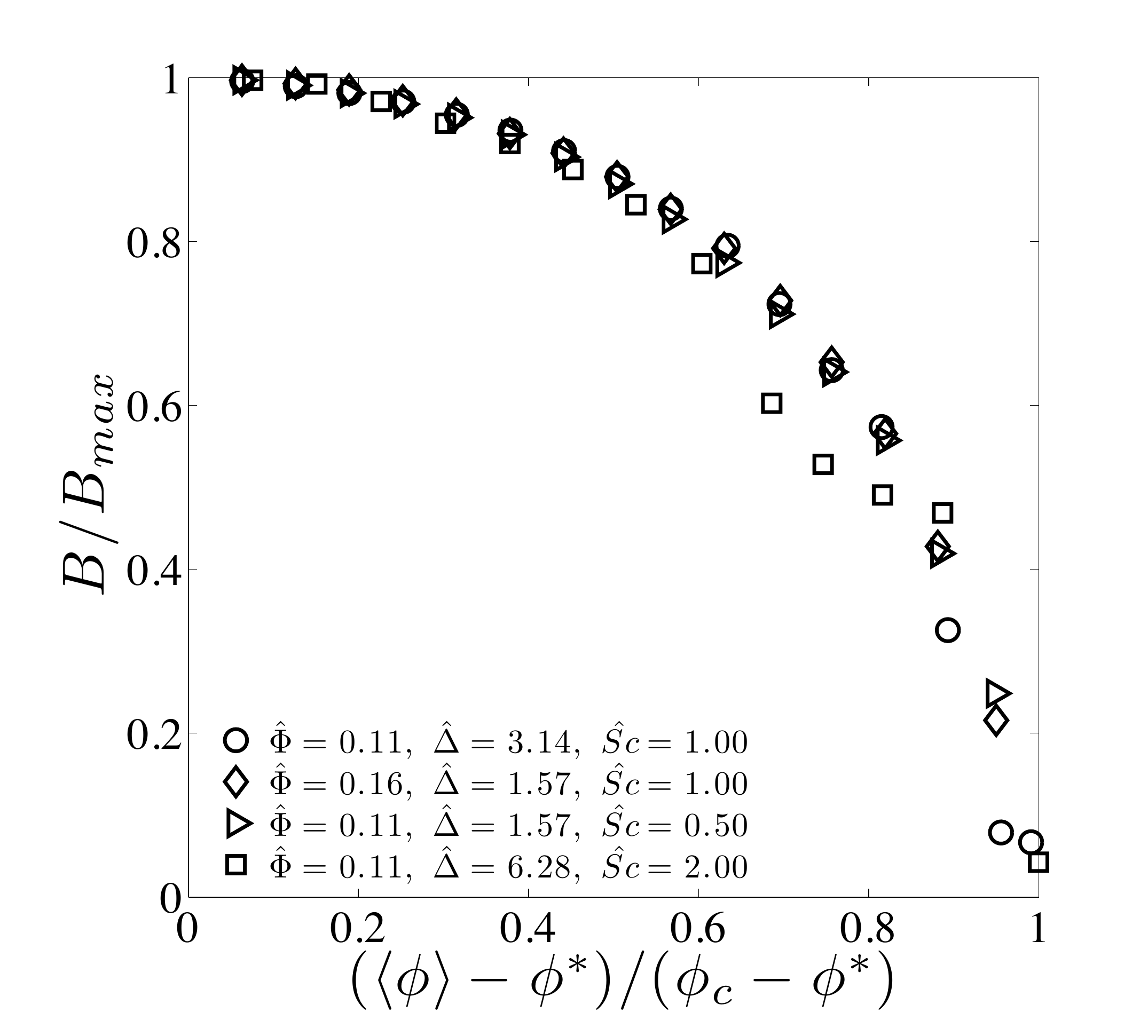} }
\caption{Dependence of $B/B_{max}$ on $\left(\left<\phi\right>-\phi^*\right)/\left( \phi_c-\phi^*\right)$ for various values (a) $\hat{\Phi}$, (b) $\hat{\Delta}$, and (c)$\hat{Sc}$.  In (d) we present $B/B_{max}$ versus $\left(\left<\phi\right>-\phi^*\right)/\left( \phi_c-\phi^*\right)$ where $\hat{\Phi}^2\hat{\Delta}/\hat{Sc}$ is constant.}
\label{fig:vol_fr_scaling_delta_right}
\end{figure}

\begin{figure}
\subfigure[]{\includegraphics[width=2.5 in]{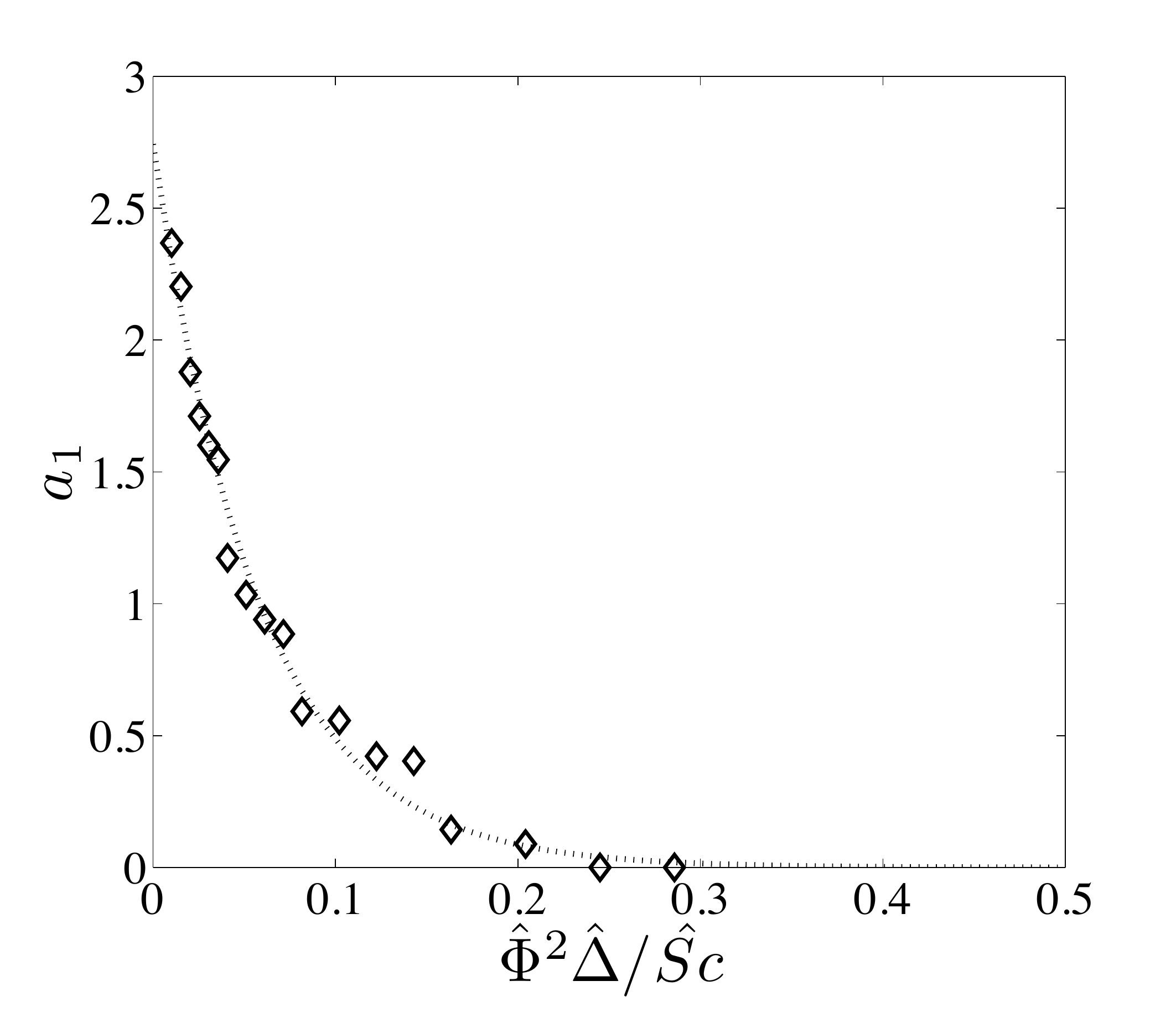} }
\subfigure[]{\includegraphics[width=2.5 in]{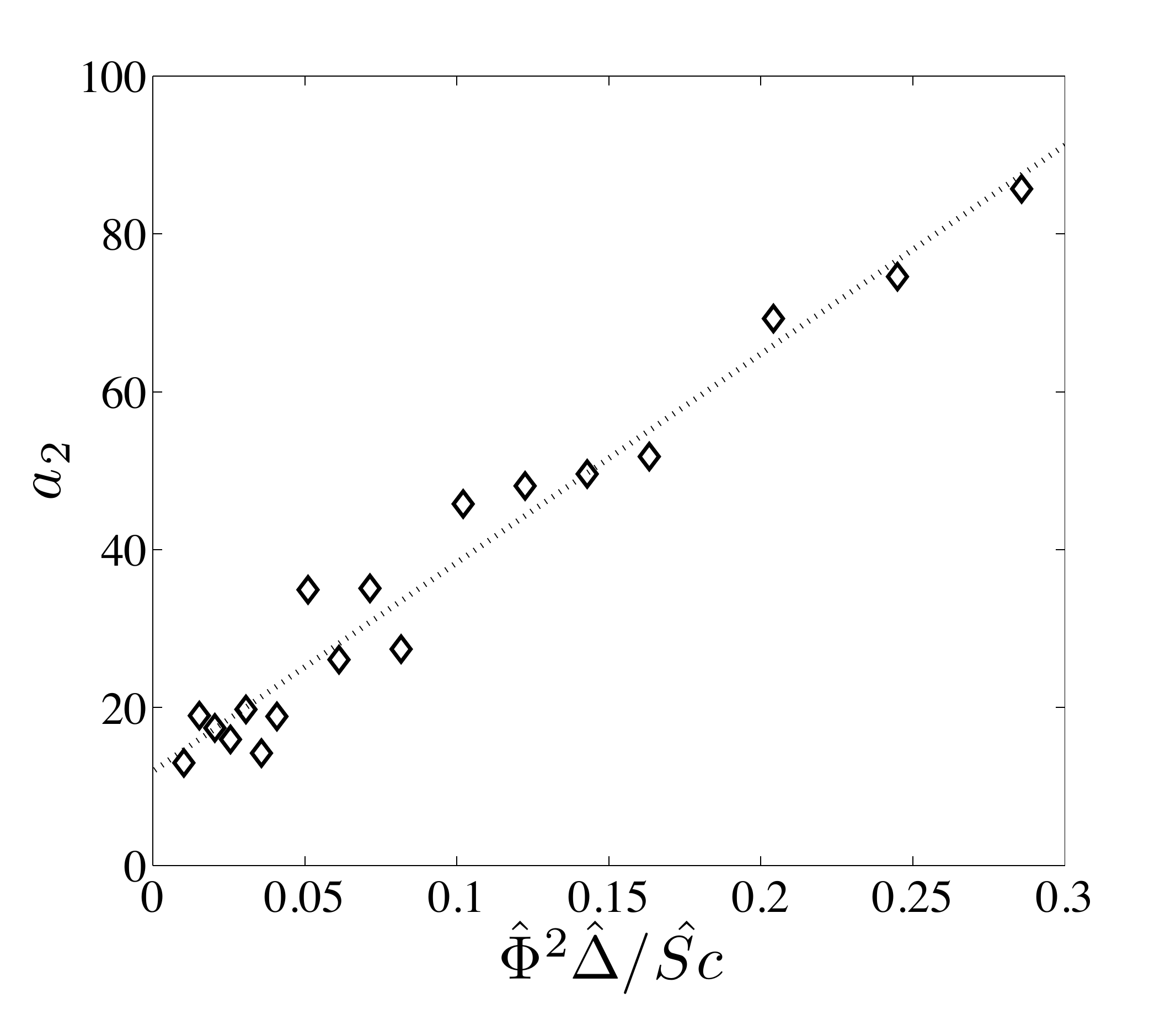} }
\caption{Fit parameters describing the dependence of $B/B_{max}$ with $\left<\phi\right>$.  (a) Gives the dependence of fit parameter $a_1$ on $\hat{\Phi}^2\hat{\Delta}/\hat{Sc}$, while (b) illustrates the behavior of fit parameter $a_2$ versus $\hat{\Phi}^2\hat{\Delta}/\hat{Sc}$.  The curve fit functions for $a_1$ and $a_2$ are given by eqs.~(\ref{eq:fit_a1}) and~(\ref{eq:fit_a2}), respectively. }
\label{fig:fit_params}
\end{figure}

Figures~\ref{fig:vol_fr_scaling_delta_right} (a)$-$(c), show the variation of $B/B_{max}$ with $\left(\left<\phi\right>-\phi^*\right)/\left( \phi_c-\phi^*\right)$ for different values of $\hat{\Phi}$, $\hat{\Delta}$, and $\hat{Sc}$.  The $\hat{Sc}$ variation in the dependence of $B/B_{max}$ on $\left(\left<\phi\right>-\phi^*\right)/\left( \phi_c-\phi^*\right)$ is clear, while the $\hat{\Phi}$ and $\hat{\Delta}$ dependence is substantially weaker.  In Figure~\ref{fig:vol_fr_scaling_delta_right} (d) we illustrate that the dependence of $B/B_{max}$ on $\left(\left<\phi\right>-\phi^*\right)/\left( \phi_c-\phi^*\right)$ for different $\hat{\Phi}$, $\hat{\Delta}$, and $\hat{Sc}$ values can be collapsed onto a single curve provided the value of $\hat{\Phi}^2\hat{\Delta}/\hat{Sc}$ is kept constant.  From Figures~\ref{fig:vol_fr_scaling_delta_left} (d) and~\ref{fig:vol_fr_scaling_delta_right} (d) it is clear that the relevant scaling parameter governing the shape of the dependence of $B/B_{max}$ on $\left<\phi\right>$ on both sides of $\phi^*$ is dictated by a single parameter given as $\hat{\Phi}^2\hat{\Delta}/\hat{Sc}$.  The following functional forms are used to describe the $\left<\phi\right>$ dependence of $B/B_{max}$: 
\begin{equation}
\frac{B}{B_{max}}=\left( \frac{1-\left(1-\left<\phi\right>/\phi^*\right)^3}{1+a_1\left(1-\left<\phi\right>/\phi^*\right)^3} \right), \quad \left<\phi\right> <\phi^*
\label{eq:fit_B_ov_B_max_left}
\end{equation} 
\begin{equation}
\frac{B}{B_{max}}=1-\left( \frac{1-\left(1-\left(\left<\phi\right>-\phi^*\right)/\left(\phi_c-\phi^*\right)\right)^2}{1+a_2\left(1-\left(\left<\phi\right>-\phi^*\right)/\left(\phi_c-\phi^*\right)\right)^2} \right), \quad \left<\phi\right> \geq\phi^*.
\label{eq:fit_B_ov_B_max_right}
\end{equation}
Here, $a_1$ and $a_2$ are least-squares fit parameters that depend only on $\hat{\Phi}^2\hat{\Delta}/\hat{Sc}$, see Figures~\ref{fig:fit_params} (a) and (b).  Both model parameters are given by smooth functions of $\hat{\Phi}^2\hat{\Delta}/\hat{Sc}$ presented in Table~\ref{tab:table3}.  It should be noted that the observed fluctuation in $a_2$ about the model curve in Figure~\ref{fig:fit_params} (b) can be attributed to the uncertainty in determining $\phi_c$ from the simulation results for $\eta_{\Delta}'$.

\begin{figure}
\subfigure[]{\includegraphics[width=2.5 in]{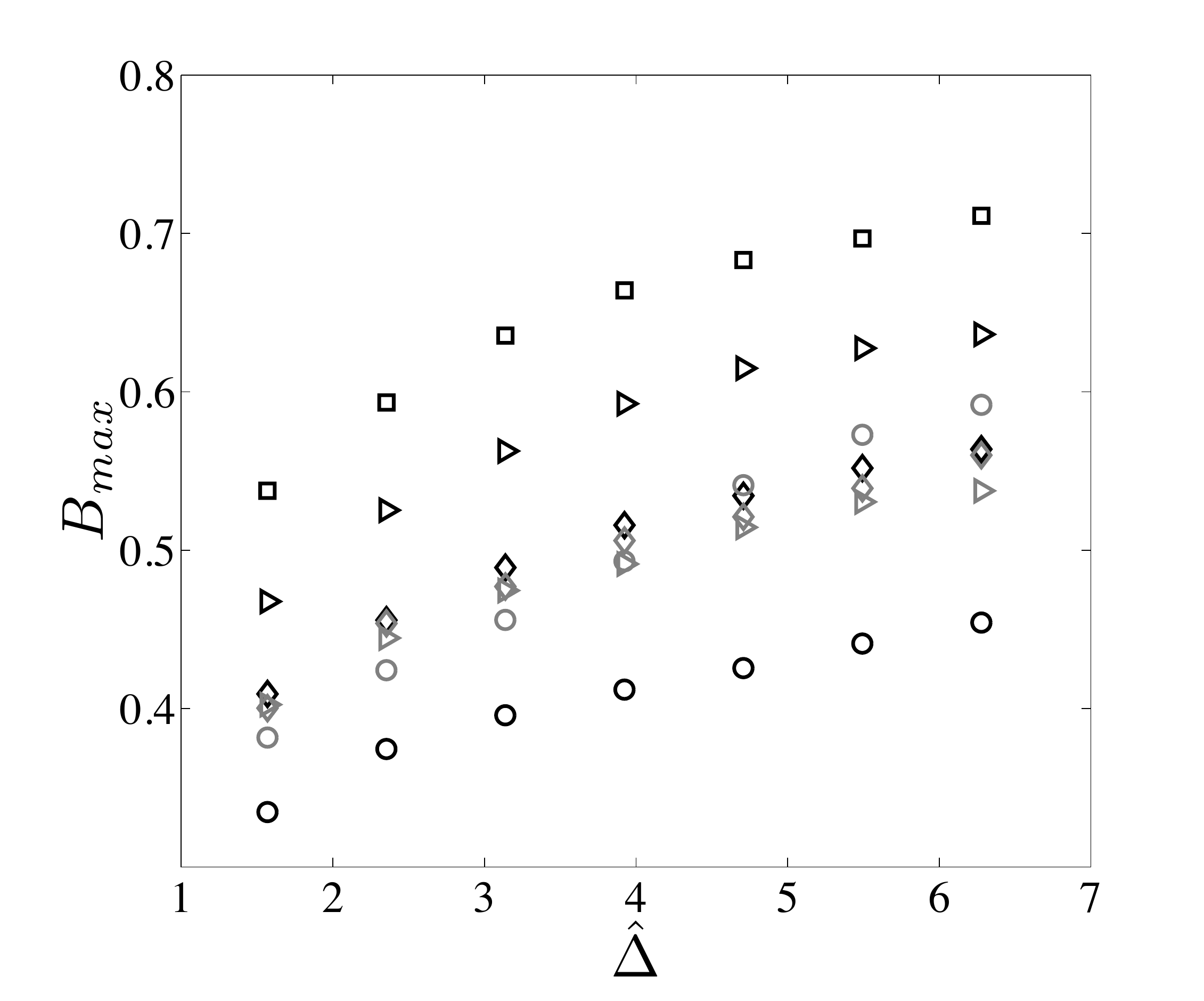} }
\subfigure[]{\includegraphics[width=2.5 in]{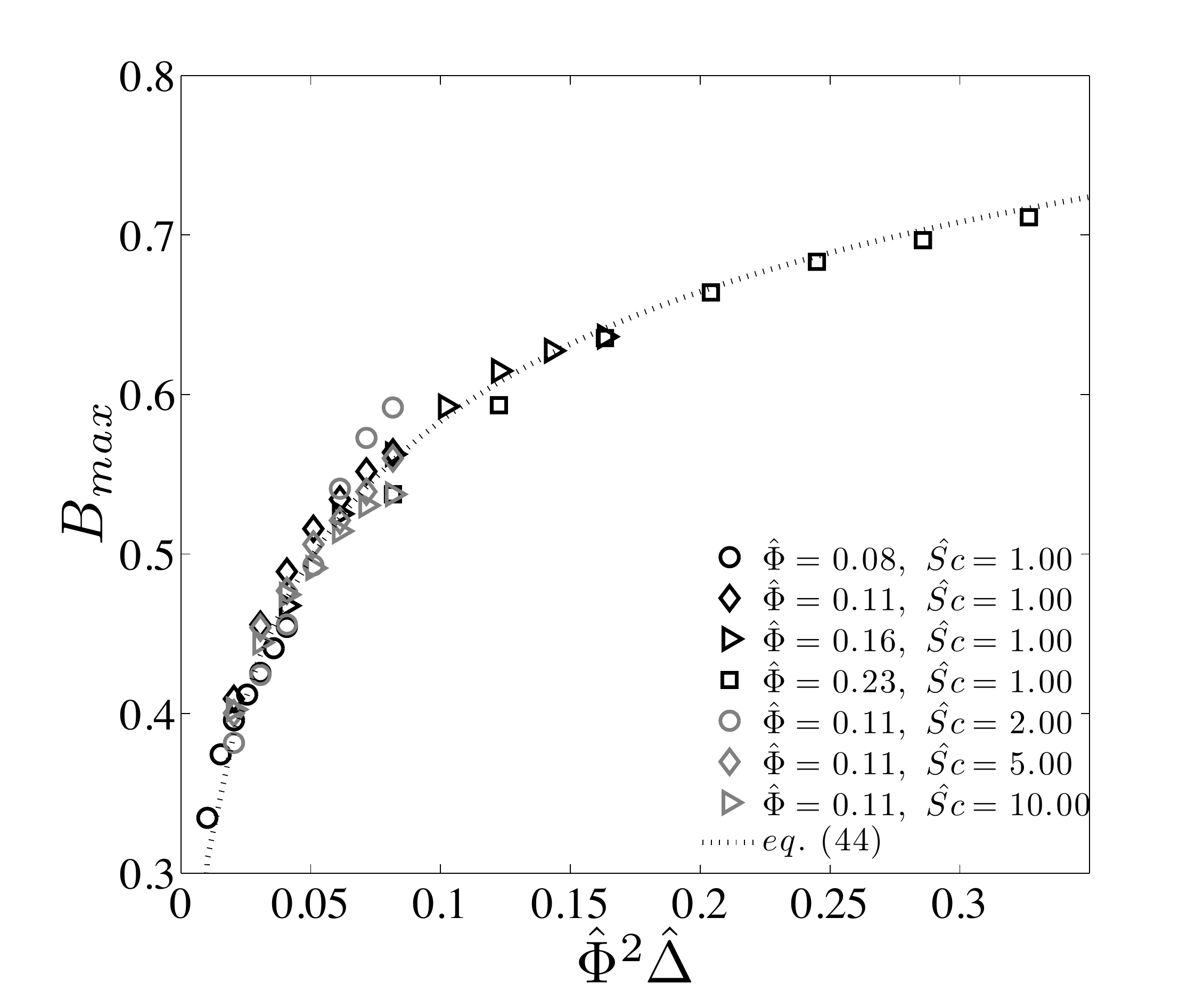} }
\caption{We explore the dependence of the extrapolated values of $B_{max}$ on (a) $\hat{\Delta}$ and (b) $\hat{\Phi}^2\hat{\Delta}$ for various values of $\hat{\Phi}$ and $\hat{Sc}$.  A clear collapse of our simulation results is obtained when plotting $B_{max}$ versus $\hat{\Phi}^2\hat{\Delta}$.}
\label{fig:B_dep_delta}
\end{figure}

In Figure~\ref{fig:B_dep_delta} (a) $B_{max}$ is presented as a function of $\hat{\Delta}$ for a variety of different $\hat{\Phi}$ and $\hat{Sc}$ values.  The variation of $B_{max}$ with $\hat{\Phi}$, $\hat{\Delta}$, and $\hat{Sc}$ is evident by inspection of Figure~\ref{fig:B_dep_delta} (a).  However, by replotting all data as a function of $\hat{\Phi}^2\hat{\Delta}$ alone one can collapse all $B_{max}$ values onto a single master curve in Figure~\ref{fig:B_dep_delta} (b).  A curve fit is presented for $B_{max}$ in terms of $\hat{\Phi}^2\hat{\Delta}$ in Table 3.

Linear extrapolation of $\phi^*$ to infinite resolution reveals that $\phi^*$ varies between $0.34<\phi^*<0.50$, with no discernable trend in $\hat{\Phi}$, $\hat{\Delta}$, or $\hat{Sc}$.  This variation in $\phi^*$ arises as a result of the fact that the \emph{cluster-scale} effectiveness factor is a weak function of $\left<\phi\right>$ in the region around $\phi^*$ and thus determining the value of the $\phi^*$ is subject to error.  Since no discernable trend in $\phi^*$ was observed, we recommend the use of $\phi^*=0.42$ because it represents the ensemble average of the different $\phi^*$ values obtained.  Moreover, since $B$ only varies slightly in the vicinity of $B_{max}$, small errors in $\phi^*$ will not influence the quantitative behavior our \emph{filtered} model substantially.  In addition, the extrapolated value of $\phi_c$ was found to vary between $0.58 < \phi_c<0.62$ with no systematic dependence on $\hat{\Phi}$, $\hat{\Delta}$, or $\hat{Sc}$.  As a result of this fluctuation we choose $\phi_c=0.59$ because this is consistent with earlier work in our group suggesting that \emph{filtered} model corrections for gas-particle hydrodynamics become negligible at volume fractions of $0.59$ and higher~\citep{Igci2011b}.

The \emph{cluster-scale} effectiveness model developed in this work predicts that effective reaction rates observed in coarse-grid simulations be larger than those predicted in fine grid simulations if the effects of fine scale structure are not accounted for via the \emph{cluster-scale} effectiveness factor.  Therefore, in order to accurately perform continuum model simulations of reacting gas-particle flows on coarse spatial grids, \emph{filtered} models for effective reactions rates are necessary.  Without these corrections, coarse-grid continuum model simulations will consistently over-estimate the conversion observed in a solid catalyzed reaction (for example see~\cite{Zimmermann2005}).  Therefore, in order to perform accurate coarse-grid numerical simulations of a first-order, isothermal, solid-catalyzed gas-phase reaction we suggest the use of the following \emph{filtered} species balance equation
\begin{equation}
\begin{aligned}
\frac{\partial \left( \rho_g \left<\epsilon_g\right>\left<\chi_{g_{A}}\right>\right)}{\partial t}+\nabla \cdot \left(\rho_g\left<\epsilon_g\right> \left<\chi_{g_{A}}\right>\left<\boldsymbol{u}\right>\right)=& -\nabla \cdot \left(D_\text{eff}\nabla\left<\chi_{g_{A}}\right>\right) -k_\text{eff}\rho_g \eta_{\Delta}' \left<\phi\right>\left<\chi_{g_{A}}\right> \\ &- k_\text{eff}\rho_g m_2\nabla\left<\phi\right>^T\cdot\nabla\left<\chi_{g_{i}}\right>,
\end{aligned}
\label{eq:filtered_model}
\end{equation}
where the constitutive relation for $\eta_{\Delta}'$ is given in Table~\ref{tab:table3}, and a constitutive relation for the effective dispersion coefficient $D_\text{eff}$ can be inferred from the work of~\cite{Loezos2002}.  In section~\ref{sec:prelim_study}, it was shown in that the non-local correction to the effectiveness factor contribute much as $37\%$ to the observed value of $\eta_{\Delta}$.  We have observed that this contribution decreases as a function of increasing grid resolution and filter size.  Therefore, in the limit of large filter sizes we expect that the contribution of the non-local term $m_2\nabla\left<\phi\right>^T\cdot\nabla\left<\chi_{g_{i}}\right>$ in eq.~(\ref{eq:filtered_model}) will be weak, and can thus be neglected.  However, for intermediate filter sizes we recommend the inclusion of this non-local correction.

\begin{table}
\begin{tabularx}{5 in}{X}
\cline{1-1}
\begin{equation}
B_{max}=\frac{4.4\sqrt{\hat{\Phi}^2\hat{\Delta}}}{1+4.4\sqrt{\hat{\Phi}^2\hat{\Delta}}}
\label{eq:B_max}
\end{equation}
\begin{equation}
\eta_{\Delta}'=\Bigg\{ \begin{array}{c c}
1-B_{max}\left( \frac{1-\left(1-\left<\phi\right>/\phi^*\right)^3}{1+a_1\left(1-\left<\phi\right>/\phi^*\right)^3}\right) &  \quad \left<\phi\right> < \phi^*  \\
 1-B_{max}\left(\frac{\left(1+a_2\right)\left(1-\left(\left<\phi\right>-\phi^*\right)/\left(\phi_c-\phi^*\right)\right)^2}{1+a_2\left(1-\left(\left<\phi\right>-\phi^*\right)/\left(\phi_c-\phi^*\right)\right)^2}\right)& \quad \phi^* \leq \left<\phi\right> < \phi_c \\
\end{array}
\end{equation}
\begin{equation}
a_1=\frac{2.76}{\exp{\left(17.2\hat{\Phi}^2\hat{\Delta}/\hat{Sc}\right)}}
\label{eq:fit_a1}
\end{equation} 
\begin{equation}
a_2=265\hat{\Phi}^2\hat{\Delta}/\hat{Sc} +10.4
\label{eq:fit_a2}
\end{equation}
\end{tabularx}
\caption{Filtered model for the \emph{cluster-scale} effectiveness factor valid for a first order, solid catalyzed, isothermal reaction.}
\label{tab:table3}
\end{table}

While in this work the evolution of species mass fraction was solved in the gas phase alone, one could envision a model for a solid-catalyzed, gas phase reaction where the species mass fraction in both gas and solid phases is tracked separately and coupled through a mass transport term between particle and fluid-phases.  In a model framework of this type, the effective interphase mass transfer rates will be decreased due to the presence of clustering in the gas-particle flow, and any reduction in reactant conversion that arises can be attributed to decreased mass transfer efficiency.  We expect that such an observed decrease in mass transfer coefficient will be on the order of the \emph{cluster-scale} effectiveness factor developed in this work.  As an example, consider a mass transfer operation taking place between FCC particles and air, where the effective mass transfer coefficient is calculated to be $k_m\approx O(0.1) \ m/sec$ using a model developed by~\cite{Gunn1978} for fixed beds.  Using $k_{m}\approx0.1 \  m/sec$ one can determine the characteristic rate constant for mass transfer $k'=k_mS_p/V_p$, where $S_p$ and $V_p$ are the surface area and volume of an FCC particle, respectively.  For an FCC particle $k'\approx 8000 \ sec^{-1}$, and using the \emph{cluster-scale} effectiveness factor model developed in eq.~(\ref{eq:B_max}) for a filter size of $2 \ cm$ we predict that $\eta_{\Delta}'=0.07$.  Using this minimum value of the \emph{cluster-scale} effectiveness factor we predict that  $k'_\text{eff}=\eta_{\Delta}'k'=530 \ sec^{-1}$ $-$ a nearly 20 fold decrease in the effective mass transfer coefficient!  Indeed decreased effective mass transfer coefficients on this order have been found in Energy Minimization Multi-Scale Model (EMMS) simulations of mass transport processes in gas-particle flows ~\citep{Dong2008}.


\section{Conclusions}
The need for the development of \emph{filtered} two-fluid models for reacting gas-particle flows is demonstrated by considering a model first-order, isothermal solid-catalyzed gas-phase reaction.  It is shown that constitutive relations for the \emph{filtered} reaction rate and \emph{filtered} species dispersion must be postulated to close the \emph{filtered} species balance equation.  Due to the fact that the model gas-phase reaction in this work is isothermal and produces no volume change, only \emph{filtered} species balance equations must be developed, without the need to alter the existing \emph{filtered} gas and solid momentum balance equations given in the earlier works of~\cite{Igci2008} and~\cite{Igci2011b} for non-reacting, monodisperse gas-particle flows.  Using fine-grid continuum model simulations of reacting gas-particle flows we extract the \emph{cluster-scale} effectiveness factor, defined as the ratio of the fine-grid reaction rate to the reaction rate observed in a coarse grid simulation.  The \emph{cluster-scale} effectiveness factor is observed to retain an inverted bell shaped dependence on volume fraction approaching unity in both the low and high volume fraction limits.  At intermediate volume fractions a decrease in the \emph{cluster-scale} effectiveness factor from unity is observed with the magnitude of this reduction being a strong function of \emph{meso-scale} Thiele modulus $\hat{\Phi}$, and dimensionless filter size $\hat{\Delta}$ with only a weak dependence on Schmidt number $\hat{Sc}$.  

Due to the sensitivity of the \emph{cluster-scale} effectiveness factor with grid size, we determine an asymptotic form for the cluster-scale effectiveness factor relying on a Richardson extrapolation of the minimum in the cluster scale effectiveness factor.  The extrapolated values of the cluster-scale effectiveness factor are shown to collapse when plotted as a function of $\hat{\Phi}^2\hat{\Delta}$, with the volume fraction dependence of the cluster-scale effectiveness factor collapsing for fixed $\hat{\Phi}^2\hat{\Delta}/\hat{Sc}$.  Table~\ref{tab:table3} presents a curve fit of our collapsed results for use in coarse-grid simulations of reacting gas-particle flows with first-order reaction kinetics.  The reduction in effective reaction rates observed in this study can be used to rationalize the over-prediction in reactant conversion that was seen in the two-fluid model simulations of~\cite{Zimmermann2005}.  Finally, we note that the filtered model developed in this work is limited to first-order reaction kinetics, and was restricted to two-dimensional periodic domain simulations.  Future work should extend such analyses to other reaction kinetics, and three-dimensional bounded domains.  However, we expect that the characteristic volume fraction, Thiele modulus, filter size, and Schmidt number dependence of $\eta_{\Delta}'$ will be qualitatively similar to the results presented here. 

\section{Notation}
\begin{flushleft}
$C_D$ -- fluid-particle drag coefficient
$a$ -- particle radius \\
$a_1$ -- fit parameter for plot of $B/B_{max}$ versus $\left<\phi\right>/\phi^*$ \\
$a_2$ -- fit parameter for plot of $B/B_{max}$ versus $\left(\left<\phi\right>-\phi^*\right)/\left(\phi_c-\phi^*\right)$ \\
$B$ -- $B=1-\eta_{\Delta}'$ \\
$Bi$ -- Biot number for mass transport  $Bi=k_m a/D$ \\
$B_{max}$ -- maximum value of $B$
$d$ -- particle diameter \\
$D^*$ -- effective diffusivity of reacting species in continuum model \\
$D$ -- molecular diffusivity of reacting species \\
$D_I$ -- \emph{intra-particle} species diffusivity \\
$D_\text{eff}$ -- \emph{meso-scale} dispersion coefficient \\
$e_p$ -- coefficient of restitution of the particle phase \\
$\boldsymbol{f_D}$ -- fluid-particle drag force experienced by the fluid \\
$\boldsymbol{g}$ -- gravitational acceleration vector \\
$g_0$ -- radial distribution function at contact \\
$G(\boldsymbol{x_0},\boldsymbol{y})$ -- weight function for filtering \\
$J_\text{coll}$ -- collisional dissipation of granular energy \\
$J_\text{vis}$ -- viscous dissipation of granular energy \\
$k$ -- intrinsic reaction rate constant \\
$k'$ -- effective rate constant for mass transfer\\
$k_\text{eff}$ -- effective reaction rate constant \\
$k_\text{eff}'$ -- \emph{meso-scale} effective rate constant for mass transfer\\
$k_m$ -- convective mass transport coefficient \\
$m_2$ -- $m_2=\Delta^2/12$ \\
$p_g$ -- gas-phase pressure \\
$\boldsymbol{q}$ -- granular energy conduction vector \\
$R_i$ -- rate of production of species \emph{i} \\
$Re_g$ -- Reynolds number for the gas-phase \\
$S_p$ -- Surface area of a particle \\
$\boldsymbol{S}$ -- rate of deformation tensor \\
$\hat{Sc}$ -- \emph{meso-scale} Schmidt number $\hat{Sc}=\mu_g/(\rho_g D)$ \\
$T$ -- granular temperature \\
$\boldsymbol{u}$ -- gas velocity \\
$\boldsymbol{v}$ -- particle velocity \\
$v_t$ -- terminal settling velocity of an isolated particle \\
$V$ -- volume of a periodic domain \\
$V_p$ -- Volume of a particle \\
$\boldsymbol{x_0}$ -- generic spatial position vector \\
$\boldsymbol{x_0}$ -- spatial position of filter center \\
$\boldsymbol{y}$ -- generic spatial position vector \\
\end{flushleft}

\begin{flushleft}
\emph{Greek letters:} \\
$\beta$ -- fluid-particle friction coefficient \\
$\Gamma_\text{slip}$ -- production of granular energy through interphase slip \\
$\delta$ -- dimensional grid size \\
$\Delta$ -- dimensional filter size \\
$\hat{\Delta}$ -- dimensionless filter size $\hat{\Delta}=\Delta |\boldsymbol{g}|/v_t^2$ \\
$\epsilon_g$ -- gas volume fraction \\
$\eta$ -- $\eta=\left(1+e_p\right)/2$ \\
$\eta_i$ -- intraparticle effectiveness factor \\
$\eta_{\Delta}$ -- \emph{cluster-scale} effectiveness factor \\
$\eta_{\Delta}'$ -- \emph{non-locally corrected} \emph{cluster-scale} effectiveness factor \\
$\kappa_i$ -- ratio of $\chi_{g_{i}}$ to its domain-averaged value \\
$\lambda$ -- conductivity of granular energy \\
$\mu_b$ -- bulk viscosity of particle phase\\
$\mu_g$ -- molecular gas-phase shear viscosity \\
$\mu_g^*$ -- effective gas-phase shear viscosity \\
$\mu_s$ -- shear viscosity of particle phase \\
$\rho_g$ -- gas density \\
$\rho_s$ -- particle density \\
$\boldsymbol{\sigma}_g$ -- gas-phase stress tensor \\
$\boldsymbol{\sigma}_s$ -- particle-phase stress tensor \\
$\phi$ -- particle volume fraction \\
$\hat{\Phi}$ -- \emph{meso-scale} Thiele modulus $\hat{\Phi}=\sqrt{k_{eff}d^2/D}$ \\
$\Phi$ -- Thiele modulus $\hat{\Phi}=\sqrt{kd^2/D_I}$ \\
$\chi_{g_{i}}$ -- mass fraction of gas species \emph{i} \\

\end{flushleft}

\section{Acknowledgements}

The authors would like to acknowledge the financial support from ExxonMobil Research $\&$ Engineering Co. and the U.S. Department of Energy, Office of Fossil Energy's Carbon Capture Simulation Initiative through the National Energy Technology Laboratory.

\appendix

\section{Effectiveness factor with \emph{non-local} contributions removed}
\label{sec:nonl_deriv}
In section~\ref{sec:filt_two_fluid} the cluster scale effectiveness factor is defined in two ways given by eqs.~(\ref{eq:effect_fac})$-$(\ref{eq:effect_fac2}).  Here, the effectiveness factor given in eq.~(\ref{eq:effect_fac2}) is derived relying on the definition of the effectiveness factor given in eq.~(\ref{eq:effect_fac}).  Let the particle phase volume fraction $\phi$ and species mass fraction $\chi_{g_{i}}$ at any point be represented as follows
\begin{equation}
\phi(\boldsymbol{y},t)=\left<\phi\right>(\boldsymbol{y},t)+\phi '(\boldsymbol{y},t) \quad \quad \chi_{g_{i}}(\boldsymbol{y},t)=\left<\chi_{g_{i}}\right>(\boldsymbol{y},t)+\chi_{g_{i}}'(\boldsymbol{y},t), 
\label{eq:decomp}
\end{equation}
where $\boldsymbol{y}$ is a spatial variable associated with \emph{filtered} and microscopic variables, respectively, neither of which are located at the filter center.  Assuming the \emph{filtered} variables can be given by smooth functions of space the \emph{filtered} value of $\phi$ and $\chi_{g_{i}}$ can be approximated via the following Taylor series
\begin{equation}
\left<\phi\right>(\boldsymbol{y},t)=\left<\phi\right>(\boldsymbol{x_0},t)+(\boldsymbol{y}-\boldsymbol{x_0})\cdot \nabla \left<\phi\right>\vline_{\left(\boldsymbol{x_0},t\right)} +O( (\boldsymbol{y}-\boldsymbol{x_0})\cdot(\boldsymbol{y}-\boldsymbol{x_0}))
\label{eq:phi_Tayl_filt}
\end{equation}
\begin{equation}
\left<\chi_{g_{i}}\right>(\boldsymbol{y},t)=\left<\chi_{g_{i}}\right>(\boldsymbol{x_0},t)+(\boldsymbol{y}-\boldsymbol{x_0})\cdot \nabla \left<\chi_{g_{i}}\right>\vline_{\left(\boldsymbol{x_0},t\right)}+O( (\boldsymbol{y}-\boldsymbol{x_0})\cdot(\boldsymbol{y}-\boldsymbol{x_0})).
\label{eq:Tayl_filt}
\end{equation}
The Taylor expansions given in eqs.~(\ref{eq:phi_Tayl_filt}) and~(\ref{eq:Tayl_filt}) can be plugged into eq.~(\ref{eq:decomp}) to yield expressions for $\phi$ and $\chi_{g_{i}}$ that depend only on the location of the filter center $\boldsymbol{x_0}$ and the microscopic spatial variable $\boldsymbol{y}$.  Utilizing eqs.~(\ref{eq:decomp})$-$(\ref{eq:Tayl_filt}) the \emph{filtered} product of $\phi$ and $\chi_{g_{i}}$ can be expressed as 
\begin{equation}
\begin{aligned}
\left<\phi \chi_{g_{i}}\right>(\boldsymbol{x_0},t)=& \left<\phi\right>(\boldsymbol{x_0},t)\left<\chi_{g_{i}}\right>(\boldsymbol{x_0},t) \\ & +m_2\nabla\left<\phi\right>^T\cdot\nabla\left<\chi_{g_{i}}\right>\vline_{\left(\boldsymbol{x_0},t\right)}
\\ &+\int_{V}\phi '(\boldsymbol{y},t)\chi_{g_{i}}'(\boldsymbol{y},t)G(\boldsymbol{x_0},\boldsymbol{y})dV
\end{aligned}
\label{eq:non_l_rem}
\end{equation}
where $m_2=\Delta^2/12$.  Removing the gradient term that appears on the right hand side of eq.~(\ref{eq:non_l_rem})  provides a method to define \emph{filtered} constitutive relations for $\eta_{\Delta}'$ that depend on local \emph{filtered} quantities alone, without considering gradients in local \emph{filtered} variables.  This approach of removing non-local effects in the process of \emph{filtered} model development has recently been advanced when considering the development of \emph{filtered} fluid-particle drag models for monodisperse gas-particle flows~\citep{parmentier2011}, and this formulation is extended here to reacting gas-particle flows.  Combining the definition of $\eta_{\Delta}$ and $\eta_{\Delta}'$ given in eqs.~(\ref{eq:effect_fac}) and~(\ref{eq:effect_fac2}) with eq.~(\ref{eq:non_l_rem}) one can arrive at the following two expressions
\begin{equation}
\begin{aligned}
\left(\eta_{\Delta}-1\right)\left<\phi\right>(\boldsymbol{x_0},t)\left<\chi_{g_{i}}\right>(\boldsymbol{x_0},t)= & \int_{V}\phi '(\boldsymbol{y},t)\chi_{g_{i}}'(\boldsymbol{y},t)G(\boldsymbol{x_0},\boldsymbol{y})dV  \\&+m_2\nabla\left<\phi\right>^T\cdot\nabla\left<\chi_{g_{i}}\right>\vline_{\left(\boldsymbol{x_0},t\right)}.
\end{aligned}
\end{equation}
\begin{equation}
\left(\eta_{\Delta}'-1\right)\left<\phi\right>(\boldsymbol{x_0},t)\left<\chi_{g_{i}}\right>(\boldsymbol{x_0},t)=\int_{V}\phi '(\boldsymbol{y},t)\chi_{g_{i}}'(\boldsymbol{y},t)G(\boldsymbol{x_0},\boldsymbol{y})dV.
\end{equation}
Therefore, $\eta_{\Delta}'$ provides a direct measure of the product of local fluctuations in $\left<\phi\right>$ and $\left<\chi_{g_{i}}\right>$, while the value of $\eta_{\Delta}$ is influenced by gradients in local filtered quantities.  In order to interrogate the effect of local fluctuations alone, we choose to model $\eta_{\Delta}'$ in this study.
\bibliographystyle{elsarticle-harv}

\begin{thebibliography}{24}
\expandafter\ifx\csname natexlab\endcsname\relax\def\natexlab#1{#1}\fi
\expandafter\ifx\csname url\endcsname\relax
  \def\url#1{\texttt{#1}}\fi
\expandafter\ifx\csname urlprefix\endcsname\relax\def\urlprefix{URL }\fi

\bibitem[{Agrawal et~al.(2001)Agrawal, Loezos, Syamlal, and
  Sundaresan}]{Agrawal2001}
Agrawal, K., Loezos, P.~N., Syamlal, M., Sundaresan, S., Oct. 2001. {The role
  of meso-scale structures in rapid gas–solid flows}. Journal of Fluid
  Mechanics 445, 151--185.

\bibitem[{{Andrews IV} et~al.(2005){Andrews IV}, Loezos, and
  Sundaresan}]{Andrews2005}
{Andrews IV}, A.~T., Loezos, P.~N., Sundaresan, S., Aug. 2005. {Coarse-Grid
  Simulation of Gas-Particle Flows in Vertical Risers}. Industrial $\&$
  Engineering Chemistry Research 44~(16), 6022--6037.

\bibitem[{Cloete et~al.(2011)Cloete, Amini, and Johansen}]{Cloete2011}
Cloete, S., Amini, S., Johansen, S.~T., Jun. 2011. {On the effect of cluster
  resolution in riser flows on momentum and reaction kinetic interaction}.
  Powder Technology 210~(1), 6--17.

\bibitem[{Dong et~al.(2008{\natexlab{a}})Dong, Wang, and Li}]{Dong2008}
Dong, W., Wang, W., Li, J., May 2008{\natexlab{a}}. {A multiscale mass transfer
  model for gas–solid riser flows: Part 1 — Sub-grid model and simple
  tests}. Chemical Engineering Science 63~(10), 2798--2810.

\bibitem[{Dong et~al.(2008{\natexlab{b}})Dong, Wang, and Li}]{Dong2008a}
Dong, W., Wang, W., Li, J., May 2008{\natexlab{b}}. {A multiscale mass transfer
  model for gas–solid riser flows: Part II—Sub-grid simulation of ozone
  decomposition}. Chemical Engineering Science 63~(10), 2811--2823.

\bibitem[{Fan and Zhu(1998)}]{Fan1998}
Fan, L.~S., Zhu, C., 1998. {Principles of gas-solid flow}. Cambridge University
  Press, Cambridge.

\bibitem[{Gidaspow(1994)}]{Gidaspow1994}
Gidaspow, D., 1994. {Multiphase Flow and Fluidization: Continuum and Kinetic
  Theory descriptions}. Academic Press Inc., San Diego.

\bibitem[{Glasser et~al.(1998)Glasser, Sundaresan, and
  Kevrekidis}]{Glasser1998}
Glasser, B., Sundaresan, S., Kevrekidis, I., Aug. 1998. {From Bubbles to
  Clusters in Fluidized Beds}. Physical Review Letters 81~(9), 1849--1852.

\bibitem[{Gunn(1978)}]{Gunn1978}
Gunn, D.~J., 1978. {Transfer of heat or mass to particles in fixed and
  fluidised beds}. Int. J. Heat Mass Transfer 21, 467--476.

\bibitem[{Igci et~al.(2008)Igci, {Andrews IV}, Sundaresan, and
  Brien}]{Igci2008}
Igci, Y., {Andrews IV}, A.~T., Sundaresan, S., Brien, T.~O., 2008. {Filtered
  Two-Fluid Models for Fluidized Gas-Particle Suspensions}. AIChE Journal
  54~(6).

\bibitem[{Igci et~al.(2011)Igci, Pannala, Benyahia, and Sundaresan}]{Igci2011a}
Igci, Y., Pannala, S., Benyahia, S., Sundaresan, S., 2011. {Validation Studies
  on Filtered Model Equations for Gas-Particle Flows in Risers}. Industrial
  $\&$ Engineering Chemistry Research, in press.

\bibitem[{Igci and Sundaresan(2011{\natexlab{a}})}]{Igci2011b}
Igci, Y., Sundaresan, S., 2011{\natexlab{a}}. {Constitutive Models for Filtered
  Two-Fluid Models of Fluidized Gas-Particle Flows}. Industrial $\&$
  Engineering Chemistry Research, in press.

\bibitem[{Igci and Sundaresan(2011{\natexlab{b}})}]{Igci2011}
Igci, Y., Sundaresan, S., 2011{\natexlab{b}}. {Verification of Filtered
  Two-Fluid Models for Gas-Particle Flows in Risers}. AIChE Journal 57~(10),
  2691--2707.

\bibitem[{Jackson(2000)}]{Jackson2000}
Jackson, R., 2000. {Dynamics of Fluidized Particles}. Cambridge University
  Press, Cambridge.

\bibitem[{Kashyap and Gidaspow(2010)}]{Kashyap2010}
Kashyap, M., Gidaspow, D., Oct. 2010. {Computation and measurements of mass
  transfer and dispersion coefficients in fluidized beds}. Powder Technology
  203~(1), 40--56.
\newline\urlprefix\url{http://linkinghub.elsevier.com/retrieve/pii/S0032591010%
001531}

\bibitem[{Kashyap and Gidaspow(2011)}]{Kashyap2011}
Kashyap, M., Gidaspow, D., 2011. {Measurements and Computation of Low Mass
  Transfer Coefficients for FCC Particles with Ozone Decomposition Reaction}.
  AIChE Journal 00~(0).

\bibitem[{Loezos and Sundaresan(2002)}]{Loezos2002}
Loezos, P.~N., Sundaresan, S., 2002. {The Role of Meso-Scale Structures on
  Dispersion in Gas-Particle Flows}. In: Circulating Fluidized Beds VII. pp.
  427--434.

\bibitem[{Parmentier et~al.(2011)Parmentier, Simonin, and
  Delsart}]{parmentier2011}
Parmentier, J.-F., Simonin, O., Delsart, O., 2011. {A Functional Subgrid Drift
  Velocity Model for Filtered Drag Prediction in Dense Fluidized Bed}. AIChE
  Journal 00~(0), 1--15.

\bibitem[{Rawlings and Ekerdt(2009)}]{Rawlings2009}
Rawlings, J.~B., Ekerdt, J.~G., 2009. {Chemical Reactor Analysis and Design
  Fundamentals}. Nob Hill.

\bibitem[{Roache(1998)}]{Roache1998}
Roache, P.~J., 1998. {Verification and Validation in Computational Science and
  Engineering}. Hermosa Publishers, Albuquerque, NM.

\bibitem[{Spalding(1980)}]{Spalding1980}
Spalding, D.~B., 1980. {Numerical computation of multiphase flow and
  heat-transfer}. In: Taylor, C. (Ed.), Recent Advances in Numerical Methods in
  Fluids. Pinbridge Press, Swansea.

\bibitem[{Syamlal(1998)}]{Syamlal1998}
Syamlal, M., 1998. {MFIX documentation:Numerical Techniques}. Tech. rep.,
  Department of Energy, National Energy Technology Laboratory.

\bibitem[{Syamlal and O'Brien(2003)}]{Syamlal2003}
Syamlal, M., O'Brien, T.~J., 2003. {Fluid Dynamic Simulation of O 3
  Decomposition in a Bubbling Fluidized Bed}. AIChE Journal 49~(11),
  2793--2801.

\bibitem[{Zimmermann and Taghipour(2005)}]{Zimmermann2005}
Zimmermann, S., Taghipour, F., Dec. 2005. {CFD Modeling of the Hydrodynamics
  and Reaction Kinetics of FCC Fluidized-Bed Reactors}. Industrial $\&$
  Engineering Chemistry Research 44~(26), 9818--9827.

\end{thebibliography}

\end{document}